\pdfoutput=1
\documentclass[11pt]{article}
\usepackage{standalone}
\usepackage{microtype}
\usepackage{amsmath,amsfonts,bbm}
\usepackage{tikz}
\usepackage{tensor}
\usepackage[T1]{fontenc}
\usepackage[utf8]{inputenc}
\usepackage{lmodern}
\usepackage{authblk}
\usepackage[disable]{todonotes}
\usepackage{cite}
\usepackage[margin=3cm]{geometry}
\usepackage{hyperref}
\usetikzlibrary{arrows,decorations.pathmorphing,decorations.markings,positioning}

\numberwithin{equation}{section}
\newcommand\nn{\ensuremath{\mathcal N}}
\newcommand\OO{\ensuremath{\mathcal O}}
\newcommand\CC{\ensuremath{\mathbb C}}
\newcommand\ZZ{\ensuremath{\mathbb Z}}
\newcommand\RR{\ensuremath{\mathbb R}}
\newcommand\I{\ensuremath{\mathbb I}}

\newcommand\qq{\ensuremath{\mathfrak q}}
\newcommand\ls{\ell_s}
\newcommand\RHS{\textsc{rhs}}
\newcommand\LHS{\textsc{lhs}}
\newcommand\VEV{\textsc{vev}}
\newcommand\email[1]{\footnote{e-mail: \href{mailto:#1}{\nolinkurl{#1}}}}
\renewcommand\d{\text{d}}
\DeclareMathOperator\tr{tr}
\DeclareMathOperator\U{U}
\DeclareMathOperator\SU{SU}
\DeclareMathOperator\diag{diag}
\DeclareMathOperator\im{Im}
\newcommand\thet[1]{\operatorname{\theta_{#1}}}

\makeatletter
\let\@fnsymbol\@arabic
\makeatother

\author{Eduardo Conde\email{econdepe@ulb.ac.be} }
\author{Micha Moskovic\email{mmoskovi@ulb.ac.be}}
\affil{
Service de Physique Th\'eorique et Math\'ematique,\\
 Universit\'e Libre de Bruxelles and International Solvay Institutes,\\
 Campus de la Plaine, CP 231, B-1050 Bruxelles, Belgium
}
\title{\bfseries\LARGE D-instanton probe and the enhançon mechanism\\ from a quiver gauge theory}
\date{}

\begin{document}
\maketitle
\begin{abstract}
  \normalsize
  We study the $\nn=2$ field theory realized by D3-branes on the $\CC^2/\ZZ_2$ orbifold. The
  dual supergravity solution exhibits a repulson singularity cured by the
  enhançon mechanism. By comparing the open and closed string descriptions of a probe D-instanton,
  we can compute the exact non-perturbative profile of the supergravity
  twisted field, which determines the supergravity background. We then show how the
  non-trivial IR physics of the field theory translates into the stringy effects that
  give rise to the enhançon mechanism and the associated excision procedure.
\end{abstract}
 
\listoftodos
\cleardoublepage
% \tableofcontents
\section{Introduction}
After the original proposal by Maldacena for a duality between $\nn=4$ Yang-Mills theory and type IIB superstrings on $AdS_5\times S^5$ \cite{Maldacena:1997re}, a lot of work focused on the construction of string theory duals to more realistic field theories.
One of the directions that proved most fruitful consists in placing D3-branes on a singular Calabi-Yau threefold in order to break supersymmetry down to $\nn=1$.
The simplest example, the conifold, was studied by Klebanov and Witten in \cite{Klebanov:1998hh}.
The low-energy dynamics of D3-branes on the conifold is described by a conformal two-node quiver gauge theory, with gauge group $\SU(N)\times\SU(N)$.
Adding $M$ fractional branes to this setup, one can engineer a theory with unequal ranks for the two factors of the gauge group, which now have non-vanishing $\beta$-functions.
The corresponding supergravity dual was found by Klebanov and Strassler in \cite{Klebanov:2000hb}.
A remarkable aspect of the solution is that the Ramond-Ramond fluxes have a logarithmic dependence on the radial coordinate, which corresponds in the field theory to a cascade of Seiberg dualities.
A second remarkable aspect is that the conifold gets deformed in the IR, corresponding to confinement in the gauge theory dual.

From the field theory point of view, a close cousin of the conifold is the $\CC^2/\ZZ_2$ orbifold: the field theory corresponding to D3-branes on this orbifold is also a two-node quiver gauge theory, which now preserves $\nn=2$ supersymmetry.
By giving appropriate mass terms to the two adjoint chiral multiplets, one can make this theory flow to the Klebanov-Witten one \cite{Klebanov:1998hh}.
For equal ranks of the gauge groups, the theory is again conformal \cite{Lawrence:1998ja} and the dual supergravity background is simply a $\ZZ_2$ orbifold of the five-sphere in the $AdS_5\times S^5$ solution \cite{Kachru:1998ys}.
Similarly to what is done in the $\nn=1$ case, one can break conformality by taking the ranks to be different; this again corresponds to adding $M$ fractional D3-branes to the $N$ regular ones.
The supergravity dual was found by \cite{Bertolini:2000dk,Polchinski:2000mx} following \cite{Klebanov:1999rd}, and presents several puzzling features.
Firstly, like for its $\nn=1$ counterpart, the logarithmic dependence of the fluxes on the radial coordinate calls for a dual which is a cascading field theory.
But Seiberg duality is a purely $\nn=1$ phenomenon, which complicated early attempts towards a field theory interpretation \cite{Aharony:2000pp,Petrini:2001fk,Billo:2001vg}.
Eventually, the authors of \cite{Benini:2008ir} put forward a consistent picture for the mechanism responsible for the cascade in analogy with the baryonic root transition of $\nn=2$ SQCD \cite{Argyres:1996eh}.
However, the main puzzle is that the supergravity solution has a singularity in the IR of the repulson type \cite{Kallosh:1995yz,Cvetic:1995mx,Behrndt:1995tr}: there is a region where a probe experiences a repulsive force, which makes the solution unphysical.
That the solution is singular could be expected on general grounds: $\nn=2$ theories do not confine and correspondingly there is no $\nn=2$-preserving deformation of the $S^5/\ZZ_2$ space that could cure the singularity as happens for the Klebanov-Strassler solution.
The singularity must be resolved differently by string theory and it was argued that in holographic duals to $\nn=2$ theories, this happens through the enhançon mechanism \cite{Johnson:1999qt}.
At a finite value of the radial coordinate, the enhançon radius, the supergravity solution cannot be trusted anymore because some branes become tensionless, providing new light degrees of freedom that are not described by supergravity and can possibly be responsible for the resolution of the singularity.
Drawing inspiration from the behavior of roots of the Seiberg-Witten curve, the authors of \cite{Johnson:1999qt} argue that, inside the enhançon radius, the supergravity solution must be excised and replaced with a solution with constant fluxes, similarly to what happens inside a conducting material in Maxwell theory.
To the extent of the authors’ knowledge, this excision procedure has never been justified in full generality from a microscopic point of view, even if partial results have been obtained by focusing on limiting cases \cite{Cremonesi:2009hq}.

The present work aims to fill this gap.
We compute directly from the field theory the profile of the twisted supergravity field $\gamma$, which encodes the backreaction of the fractional branes and completely determines the supergravity solution once the configuration of regular branes is given.
This computation will be done with arbitrary values for the gauge theory couplings, which translates in the string theory dual to having arbitrary values for the string coupling $g_s$ and the string length $\ls=\sqrt{2\pi\alpha'}$, and for any point on the Coulomb branch of the theory.
We will prove that the twisted supergravity field $\gamma$ can be written in terms of field theory data as
\begin{equation}
  2\pi i\,\gamma(z) = 2\pi i\,\gamma^{(0)}  -\beta\,\int_1^{T_r(z)}
\frac{\mathrm{d}v}{\sqrt{(v^2-\alpha_1^2)(v^2-\alpha_2^2)}}\,,
  \label{eqn:resultgamma}
\end{equation}
where $z$ is a complex coordinate on the orbifold fixed plane, $\gamma^{(0)}$ is the asymptotic value of $\gamma$, $T_r$ is a ratio of polynomials encoding the choice of Coulomb branch vacuum and $\alpha_i$, $\beta$ are specific coupling-dependent constants.
All these quantities will be defined precisely in due course.
Choosing the particular vacua that have been studied from the supergravity side and taking the large $N$ limit of \eqref{eqn:resultgamma}, we can then derive that an enhançon mechanism takes place at a radius that perfectly matches the supergravity expectations: $\gamma$ is constant inside this radius, confirming the proposal of \cite{Johnson:1999qt}.

To make the proof of \eqref{eqn:resultgamma} possible, we draw on recent developments in two very different research lines.
The first of these is the use of D-brane probes to derive holographic string theory backgrounds from the field theory side \cite{Ferrari:2012nw,Ferrari:2013pi,Ferrari:2013pq,Ferrari:2013hg,Ferrari:2013wla,Ferrari:2013aza,Ferrari:2013aba,Ferrari:2013waa}.
The general starting point of this approach is a D-brane configuration that contains not only the large number of ``background'' D-branes whose near-horizon geometry corresponds to the sought-for string theory dual, but also a small number of D-branes that act as a probe.
This system then contains, in addition to the strings with both endpoints on the background branes, also strings with both ends on the probe branes and strings with mixed boundary conditions.
The crucial idea of \cite{Ferrari:2012nw} is to integrate out the background/background and probe/background open strings, which yields an effective action for the probe/probe open strings that can be interpreted as describing a probe brane in the holographically dual closed-string background.
By matching the specific action one obtains in this way with the probe action in an arbitrary supergravity background, one can then read-off the background.
Like in \cite{Ferrari:2012nw, Ferrari:2013pq}, we consider a setup where the background branes are D3-branes and the probe is a small number of D-instantons (i.e.\ D(-1)-branes).
More specifically, the probe we will use is a single fractional D(-1)-brane.
The open-string realization of the D(-1)/D3 system in flat space is known explicitly for small $\alpha'$ \cite{Green:2000ke,Billo:2002hm} and can be straightforwardly generalized to the orbifold setting \cite{Argurio:2007vqa,Argurio:2008jm,orbifold} by following the same procedure as for D3-branes \cite{Lawrence:1998ja}.
The action for D(-1)-branes in the presence of D3-branes (in a ``near-horizon'' limit) has a purely field theoretic intepretation\footnote{In this paper, we will deal only with a fractional D(-1)-brane that sits on a quiver node also occupied by D3-branes and can directly be interpreted as a gauge theory instanton. In the case where the node is occupied by at most one D3-brane, that gauge group does not receive instanton corrections in field theory, and the D(-1)-brane corresponds to a ``stringy instanton''. Nevertheless, it turns out that also stringy instantons can be given a gauge theory interpretation in a suitable UV completion \cite{Argurio:2012iw}.} as the ADHM action for supersymmetric instantons \cite{Dorey:2002ik}.
The D(-1)-brane couples to the D3-branes through moduli that transform in an (anti-)fundamental representation of the four-dimensional gauge group and one can always integrate them out exactly.
However, the integration of the D3-brane fields involves the computation of a full-fledged non-chiral correlator in the four-dimensional gauge theory, which seems intractable in general.
In the conformal case, this correlator turns out to be trivial and one can recover the full supergravity background by matching the action for several D(-1)-branes with the non-Abelian probe brane action of \cite{Myers:1999ps,Taylor:1999pr}.
This is also the case for regular D3-branes on orbifold singularities whose field theory description is conformal.
One can then generalize the construction of \cite{Ferrari:2012nw}, taking also the D(-1)-branes to be regular, and recover the corresponding string theory duals \cite{orbifold}.
Unfortunately, for the $\CC^2/\ZZ_2$ orbifold with fractional D3-branes which is the focus of the present paper, the theory is non-conformal and one cannot reconstruct the full supergravity multiplet in this way.
We circumvent this difficulty by using a fractional D(-1)-brane as a probe instead of a regular one. 
This brane couples only to the twisted sector at the orbifold singularity, which captures the essential information on the background.
Applying this procedure in this case yields the following identity relating the twisted supergravity field $\gamma$ to a field theory correlator,
\begin{equation}
\gamma(z)=\gamma^{(0)}+\frac{i}{\pi}\left\langle\tr_{M}\log\left({z-Z_1}\right)\right\rangle
-\frac{i}{\pi}\left\langle\tr_{M}\log\left({z-Z_0}\right)\right\rangle\,,
\label{eqn:resultcorr}
\end{equation}
where $Z_0$ and $Z_1$ are the adjoint scalars of the two gauge groups normalized to have units of length.
This identity was derived in \cite{Billo:2012st} by computing string worldsheet diagrams, but we will rederive it much more straightforwardly.
The identity \eqref{eqn:resultcorr} also involves expectation values in the full gauge theory on the D3-branes and one might think naively that not much has been gained by focusing on the twisted sector.
There is a crucial difference however between these correlators and the correlator one is faced with in the untwisted sector: in \eqref{eqn:resultcorr}, only the chiral fields $Z_0$ and $Z_1$ enter.
This gives us more control and allows us to compute them explicitly by exploiting the impressive recent progress in the resummation of instanton corrections to $\nn=2$ quiver gauge theories \cite{Nekrasov:2012xe,Fucito:2012xc}.

The plan of the paper is as follows.
In section~\ref{sec:review}, we review the supergravity background corresponding to D3-branes on the $\CC^2/\ZZ_2$ orbifold and explain the enhançon mechanism that has been conjectured to cure the IR singularity.
In section~\ref{sec:micro}, we detail the microscopic model we start with, consisting of $N$ regular D3-branes, $M$ fractional D3-branes of each type and one fractional D(-1)-brane.
In section~\ref{sec:gamma}, we derive equation \eqref{eqn:resultgamma} by building the effective action for the D(-1)-brane and comparing it with the supergravity probe action.
The computation of the correlators in \eqref{eqn:resultcorr} is quite technical and we have chosen to present it separately in appendix~\ref{sec:appendix}.
In section~\ref{sec:enhancon}, we take the large $N$ limit of this result, showing explicitly that the enhançon mechanism takes place.
Finally, we conclude in section~\ref{sec:conc} by giving some perspectives on possible future work.

\section{D3-branes on the $\CC^2/\ZZ_2$ orbifold. A review}
\label{sec:review}

The $\CC^2/\ZZ_2$ orbifold is a representative of a larger family, the ADE
 orbifolds. These are built as $\CC^2/\Gamma_{\textrm{ADE}}$,
with $\Gamma_{\textrm{ADE}}$ being a discrete subgroup of $SU(2)$. The theories living on
D3-branes placed on these orbifolds are $\nn=2$ superconformal quiver gauge theories. The
Coulomb phase of these theories is non-conformal, and can be engineered in the string
picture by including fractional D3-branes. The model we are interested in, with $\Gamma_{\textrm{ADE}}=\ZZ_2$,
is also known as the affine $A_1$ quiver theory. In this section, we review what we have learnt about the workings of the
gauge/string duality in this example. Most of what we say can be found in \cite{Benini:2008ir},
where this model was thoroughly studied.

\subsection{A supergravity perspective}

Our setup is made up of a large number of parallel $N$ regular and $2M$ fractional D3-branes in\footnote{As we will later deal with instantons, it is more convenient to rotate to Euclidean signature from the start.}
$\RR^{4}\times\CC\times\CC^2/\ZZ_2$. We use coordinates $\left(x^{\mu},z,z^2,z^3\right)$
for this space, and the $\ZZ_2$ acts as $\left(z^2,z^3\right)\to\left(-z^2,-z^3\right)$.
The regular branes can probe the full transverse space $\CC\times\CC^2/\ZZ_2$,
while the fractional branes are constrained to live at the orbifold singularity,
which is the complex $z$-plane at the origin of $\CC^2$ in this case. There are two types of fractional branes, which
we will denote as type 0 and type 1, and we will consider $M$ branes of the first type,
and $M$ of the second type. A regular brane can be thought of as a bound
state of a type 0 and a type 1 fractional brane. For some purposes, it is useful to think of
the fractional D3-branes as wrapped D5-branes. Recall that the orbifold $\CC^2/\ZZ_2$ can
be seen as the singular limit of a smooth ALE manifold (in our case it is the Eguchi-Hanson
space \cite{Eguchi:1978gw}) where a homologically non-trivial 2-cycle $\Sigma$ collapses.
The type 1 and type 0 fractional D3-branes correspond to D5-branes wrapped on $\Sigma$ and
$-\Sigma$ respectively, stabilized by certain background fluxes.

The presence of fractional branes induces the excitation of some of the twisted modes of
type IIB string theory. Thinking of the fractional D3-branes as wrapped D5-branes, it is
easy to understand that the reduction of the potentials $C_2$ and $B_2$ on the exceptional
cycle $\Sigma$ will give rise to non-zero twisted scalars $c$ and $b$. These two fields can only
depend on $z,\bar z$, as the fractional D3-branes can only probe this plane, and
are conveniently combined to form the complex field:
\begin{equation}
\gamma= c+\left(C_0+i\,e^{-\Phi}\right)b=\frac{1}{2\pi\ls^2}\int_{\Sigma}\left(C_2+\frac{i}{g_s}\,B_2\right)\,,
\label{eqn:g.def}
\end{equation}
to which we will generically refer as the twisted supergravity field. In writing the last equality
we have taken into account that the axio-dilaton is constant,
$C_0+i\,e^{-\Phi}=\frac{i}{g_s}$, since it does not couple to D3-branes.
Such branes do source a $C_4$ potential, and of course backreact on the metric. Instead of
writing the expression for all these fields, which can be found for instance in
\cite{Bertolini:2000dk}, the point we want to emphasize here is that \emph{the full
type IIB background follows\footnote{Essentially the metric and the RR potential $C_4$ are
determined by a warp factor $H(z,z^1,z^2)$, which is determined itself by solving a Poisson
equation sourced by the the regular and the fractional D3-branes. The contribution of the latter comes with
a $|\partial_z\gamma|^2$ factor. The position of the former must be specified as the only extra input.} once the twisted supergravity field $\gamma$ is known}.
Because of $\nn=2$ supersymmetry, $\gamma$ depends holomorphically on $z$, i.e.\
$\partial_{\bar z}\gamma=0$.

The profile of the twisted supergravity field is in turn solely determined by the positions
of the fractional D3-branes. This follows from its equation of motion, that can be derived
from the type IIB supergravity action taking into account the twisted supergravity supermultiplet and the fractional D-brane sources:
\begin{equation}
\Delta\gamma=2i\sum_{j=1}^M\left(\delta^2(z-z_j)-\delta^2(z-\tilde{z}_j)\right)\,,
\label{eqn:g.eq}
\end{equation}
where the fractional branes of type 1 sit at positions $z_j$, and those of type 0 sit
at $\tilde{z}_j$. Notice that the profile of $\gamma$ is only sensitive to genuine fractional
D3-branes: if $z_i=\tilde z_j$ for some pair $(i,j)$, these two fractional branes form a regular
D3-brane and do not source $\gamma$ anymore, in agreement with the fact that $\gamma$ does not couple to regular branes.
It is easy to solve the two-dimensional Laplace equation \eqref{eqn:g.eq} to obtain:
\begin{equation}
\gamma=\frac{i}{\pi}\left(\sum_{j=1}^M\log(z-z_j)-\sum_{j=1}^M\log(z-\tilde{z}_j)\right)+\gamma^{(0)}\,.
\label{eqn:g.pert}
\end{equation}
The value of $\gamma^{(0)}$ is clearly the asymptotic value, as $z\to\infty$, of $\gamma$.
There is a preferred value of $b$ and $c$ for perturbative string theory:
 if we choose
$\gamma^{(0)}=\frac{i}{2g_s}\Leftrightarrow\lim_{z\to\infty}(c,b)=(0,\frac{1}{2})$,
the world-sheet propagating on this orbifold is a free CFT \cite{Aspinwall:1996mn,Blum:1997fw}.
We will see shortly that this value is also special from the field theory point of view, and we will often make this choice for simplicity.

The take-home message is then that the supergravity background is determined by the way in
which we distribute the fractional D3-branes in the geometry, and this information is encoded
in the twisted field $\gamma$. The distribution of branes is naturally related to
the different vacua of the dual gauge theory, as we now explain.

\subsection{A field theory perspective}

When we look at our brane system from far away, that is at large $|z|$, we essentially see
a stack of $N+M$ regular D3-branes on the $Z_2$ orbifold, since the fact that the positions
of type 0 and type 1 fractional are \emph{a priori} different becomes irrelevant.
We effectively obtain a theory with only $N+M$ regular branes. The field theory dual to this setup is well-known
\cite{Kachru:1998ys}. It is the $\nn=2$ superconformal quiver theory with gauge group $\SU(N+M)_0\times \SU(N+M)_1$.
This theory has a rich moduli space of vacua, with both Coulomb and Higgs branches.
The Higgs branch corresponds to giving \VEV s to the bifundamentals of the quiver. 
As is well known, the field theory on the Higgs branch is not very interesting, since both the superpotential and the K\"ahler potential are not renormalized \cite{Argyres:1996eh}.
In the brane picture, (the mesonic part of) this branch has a nice geometrical interpretation:
it corresponds to the possible configurations of regular D3-branes occupying certain positions
in the transverse space $\CC\times\CC^2/\ZZ_2$. 
Notice that in the covering space $\CC\times\CC^2$, the branes have to be arranged in pairs of orbifold images $(z,\pm z^2,\pm z^3)$.
Another possibility is to have some D3-branes at the origin of $\CC^2$ which maps to the orbifold singularity of $\CC\times \CC^2/\ZZ_2$.
Those D3-branes do not need to be paired and can have arbitrary positions along the $\CC$ plane with coordinate~$z$; those are fractional branes.
The different configurations for fractional branes correspond in the field theory to the Coulomb branch, obtained by giving expectation values to the two adjoint fields. Denoting these
fields by $\varphi_0$, $\varphi_1$ (see figure~\ref{fig:Z2fracquiver}), at the perturbative level we
can identify their respective non-zero eigenvalues with the $\tilde{z}_j$, $z_j$ of \eqref{eqn:g.eq}.
There are also mixed branches, where both bifundamentals and adjoints acquire \VEV s.

We are interested in the IR physics of the Coulomb branch. More precisely we will be mainly
concerned with the point that was dubbed ``enhançon vacuum'' in \cite{Benini:2008ir}. It
is classically defined by $\varphi_0$ having $M$ prescribed non-zero
eigenvalues, or equivalently by having $M$ fractional branes of type 0 sitting at the roots of
$\tilde{z}_j^M=-z_0^M$, where $|z_0|$ is an arbitrary UV scale. Below this scale, we are
left with an effective theory describing $N$ regular branes plus $M$ fractional branes of type 1 sitting at $z=0$. The gauge group is Higgsed
down to $SU(N)_0\times SU(N+M)_1$ if we take into account that all the $U(1)$ factors are IR free
and decouple. 
Such an effective theory is not conformal, as 
reflected by the running of $\gamma$ in equation \eqref{eqn:g.pert}, which for this vacuum
reads
\begin{equation}
\gamma=\frac{i}{\pi}\log\frac{z^M}{z^M+z_0^M}+\frac{i}{2g_s}\qquad\underset{\textrm{large }M}{\rightsquigarrow}
\qquad\gamma\approx\begin{cases}
\frac{iM}{\pi}\log\frac{z}{z_0\,e^{-\frac{\pi}{2g_sM}}} & \textrm{if }|z|<|z_0|\\
\frac{i}{2g_s} & \textrm{if }|z|>|z_0|\end{cases}
\label{eqn:g.enh}
\end{equation}
In order for the classical supergravity solution that follows from \eqref{eqn:g.enh} to be a good description of the
gauge theory, one should require as usual that $N$ and $M$ be large. Using the complexified gauge couplings
$\tau_a=\frac{\vartheta_a}{2\pi}+\frac{4\pi\,i}{g_a^2}$ for the two $\SU(N+M)_a$ factors, the holographic relations between the gauge couplings and the supergravity fields read
\begin{equation}
\tau_0+\tau_1=\frac{i}{g_s} \, , \qquad\qquad \tau_1=\gamma \, .
\label{eqn:holotaugs}
\end{equation}
The first relation is the standard holographic dictionary applied to the diagonal $\SU(N+M)$ gauge group; the second one can be shown by a fractional probe brane analysis \cite{Bertolini:2000dk}.
This implies the following relation between the bare gauge couplings and the asymptotic values of the dilaton and $\gamma$:
\begin{equation}
  8\pi g_s=g_a^2\,,\qquad \lambda_a=(N+M)g_a^2=8\pi g_s(N+M)\,,
  \label{eqn:holo1}
\end{equation}
defining the 't Hooft couplings $\lambda_a$.
In particular, we see that $g_0=g_1$, which can be traced back to the fact that we have chosen the special asymptotic value $\frac{i}{2 g_s}$ for the twisted supergravity field $\gamma$.
Notice that the relations \eqref{eqn:holotaugs} are not restricted to $z\to\infty$. Indeed the second 
one provides an exact match between the supergravity
running of $\gamma$ and the perturbative running of the gauge couplings (which is exhausted at one-loop).

Strictly speaking, the supergravity description is a faithful one for small $g_s$, large $N$ and $M$ so that
$g_s(N+M)\gg1$.
We take $g_s(N+M)$ to be a large, but finite, number.
Since below $|z_0|$ the gauge couplings run in opposite directions, at a certain scale, one of the gauge couplings blows up.
The supergravity approximation breaks down there, the second relation in \eqref{eqn:holotaugs} no longer holds, and a stringy resolution is needed.
There are several ways to proceed, related to different non-perturbative completions of the same perturbative physics.
Let us discuss this in a bit more detail below.

\subsection{Non-perturbative physics and the enhançon}
\label{ssec:enh}

With the amount of supersymmetry that we have, the perturbative series for
correlation functions of protected operators in the field theory truncate at one loop. Any other quantum correction must come from
instantons, i.e.\ with a pre-factor $e^{-l/g_a^2}$ ($l$ being a positive number). At large $N,M$ and
fixed 't Hooft coupling, $g_a^2\sim 1/(N+M)\to0$ and these corrections are exponentially suppressed.
This is why supergravity outside the enhançon matches exactly the one-loop field theory, although they are expected to be valid in opposite regimes of $\lambda_a$.
Nevertheless, it is known that non-perturbative corrections
can still contribute in the 't Hooft limit \cite{Douglas:1995nw}, as will occur in our model.
Such corrections are proportional to $e^{-l/\lambda_a}$, which does not have to be small.

Let us now follow the holographic RG flow of our theory from equation \eqref{eqn:g.enh}, assuming
that both $N$ and $M$ are large. We have a conformal theory above the scale
$|z_0|$. Below this scale, $\gamma$ starts to run, inducing a running of the couplings. Recall from
\eqref{eqn:g.def} that the imaginary part of $\gamma$ gives us the gauge coupling $1/g_1^2$ in field theory and the scalar $b$ in supergravity.
This scalar should be in the range $\left[0,1\right]$ in order to have a proper field theory interpretation with positive $g_1^2,g_0^2$.
When we reach the scale
\begin{equation}
\rho_1=|z_0|\,e^{-\frac{\pi}{2g_sM}}\,,
\label{eqn:rho_1}
\end{equation}
$\gamma$ vanishes, and so does $b$.
At this point, from equation \eqref{eqn:holotaugs}, we see that $\lambda_1$ diverges.
Past this point, we can no longer trust the supergravity solution \eqref{eqn:g.enh}.
A way to think about it is that probe fractional branes become tensionless at $\rho_1$ (the tension of such branes is proportional to $b$).
Potentially, a whole fauna of stringy phenomena, not captured by the supergravity approximation,
could arise.

Nothing dramatic happens for the supergravity solution at $\rho_1$ though, so one could think of pushing the
gauge/string duality and come up with a possible field-theoretic interpretation below this scale. This is what the
authors in \cite{Benini:2008ir} did. They proposed an interpretation of the solution for $|z|<\rho_1$
\emph{\`a la} Klebanov-Strassler: we must perform a Higgsing in the field theory, interpreted
as a large gauge transformation in the supergravity background\footnote{Notice that a large gauge
transformation is \emph{not} a gauge transformation. With it, we are changing the vacuum in the underlying field theory.
The fact that we have to perform this operation is not encoded in the supergravity background,
but it must be done by hand instead.} \cite{Benini:2007gx}.
This Higgsing is a strong coupling effect: it arises at a scale $\sim e^{-l/\lambda_a}$ where a gauge coupling blows up.
The non-trivial field theory vacuum responsible for the Higgsing is very similar to the baryonic root in $\nn=2$ SQCD \cite{Argyres:1996eh}, it has hence been called a baryonic root transition.
The rank
of the gauge group with diverging coupling is reduced by $2M$ and the beta functions flip sign. The large gauge transformation shifts the twisted field of \eqref{eqn:g.enh}
in this region as $\gamma\to\gamma+\frac{i}{g_s}$.
If we keep going down the flow, we will hit another point where $g_0$ diverges, and the same operation
must be performed on the other gauge group. This can happen multiple times:  we say that the theory cascades. Apart from the
fact that here the Higgsings are not associated to Seiberg dualites since we have $\nn=2$ supersymmetry, there is a fundamental
difference with the Klebanov-Strassler case. In the latter, at the end of the cascade, the theory confines
(its dual counterpart is the deformation of the conifold). However, our $\nn=2$ model is not confining.
A different, but very interesting, phenomenon occurs. It has come to
be known as the enhançon, as originally named in \cite{Johnson:1999qt}. Let us discuss it from both sides
of the gauge/string duality.

From the supergravity point of view, we find that the background presents a singularity (where the metric
blows up) of a peculiar type: a repulson \cite{Kallosh:1995yz,Cvetic:1995mx,Behrndt:1995tr}. Close to it, there is a region of ``anti-gravity'',
characterized by a positive sign of $\partial_{|z|}g_{00}$. From the warp
factor of the supergravity solution, one can find that this anti-gravity region starts at around the scale
\begin{equation}
\rho_e= e^{-\frac{\pi N}{g_sM^2}}\rho_1\,.
\label{eqn:rho_e}
\end{equation}
Probe branes feel a repulsive potential below $\rho_e$, and cannot enter this region. This is supported
by a computation of the D3-brane Page charge, which gives \cite{Benini:2008ir}:
\begin{equation}
\int \left(F_5+B_2\wedge F_3\right)\propto\left(N+M\left[\frac{g_sM}{\pi}\log\frac{|z|}{\rho_1}\right]\right)\,,
\label{eqn:F5}
\end{equation}
where were are denoting by $[\cdot]$ the floor function. This shows that inside the region of
radius $\rho_e$ there is an unphysical negative D3 charge.
Even if we want to believe in the supergravity solution below the scale $\rho_1$, we can only trust it
down to the smaller scale $\rho_e$. The latter is the enhançon scale.
The standard lore in supergravity is that the $M$ fractional branes that were supposed to be at the origin expanded
to form a dense ring at the enhançon scale. Since inside this ring no branes are left, we should solve \eqref{eqn:g.eq} again
with this assumption. This obviously yields a constant $\gamma$ in this region. 
This correction \emph{by hand} of $\gamma$ is commonly known as the excision procedure.
Notice that since it is done manually, we could have chosen to perform the excision procedure already at the scale $\rho_1$, or any other scale in between where one of the gauge couplings diverges.
Different choices of where to perform the excision correspond to different choices of vacua (which only differ
non-perturbatively) in the field theory. Following the terminology of \cite{Benini:2008ir}, excising at $\rho_1$ ($\rho_e$)
corresponds to the enhançon (cascading) vacuum.

There is a field-theoretical phenomenon that takes place in the large $N$ limit of $\nn=2$ gauge theories, which
resembles very much the repulson singularity we just described. It was first noticed in \cite{Johnson:1999qt}
and is called enhançon mechanism for historical reasons (having to do with enhanced symmetries). Take for example $\nn=2$ SQCD with gauge group $\SU(N)$. The IR physics is controlled by the Seiberg-Witten (SW) curve
\begin{equation}
y^2=\prod_{k=1}^N(x-\varphi_k)^2+4\Lambda^{2N}\,,
\label{eqn:SW}
\end{equation}
where $\Lambda$ is the strong coupling scale, and the $\varphi_k$ parameterize a point in the moduli space. At large $N$, for points
with $|\varphi_k|\gg\Lambda$, the branch cuts of $y(x)$ are very small and they are located near the classical
values $x=\varphi_k$. On the contrary, when $|\varphi_k|/\Lambda\to0$, the branch cuts become longer and remain at a finite distance
from the origin of the $x$-plane. They pile up at a ring of radius $2^{\frac{1}{N}}\Lambda$. If we consider the configuration
\begin{equation}
y^2=x^{2N-2}(x-\varphi)+4\Lambda^{2N}\,,
\label{eqn:SW.probe}
\end{equation}
that corresponds to a breaking $\SU(N)\to\SU(N-1)\times\U(1)$, and we track the two branch points associated to $\varphi$,
we see that for $|\varphi|\gg\Lambda$, they are close together and around $x=\varphi$. When $|\varphi|$ approaches $\Lambda$,
the branch points separate from each other, and melt into the ring of quantum roots. The branch points can never
penetrate inside this ring. Associating branch cuts with branes, clearly this resembles the enhançon phenomenon
found in supergravity.

However, to our knowledge, the connection between the SW curve physics and the enhançon mechanism
has never been established in the literature in a completely top-down approach. This is of course a difficult problem, since its solution would involve
computing non-perturbative corrections to the supergravity background. Two important steps forward in this direction
have been taken, first by Cremonesi in \cite{Cremonesi:2009hq}, and more recently by the authors of \cite{Billo:2012st} (see also \cite{Martucci:2012jk}).
The former cleverly used the M-theory uplift of a brane configuration \cite{Witten:1997sc} corresponding to pure 
$\nn=2$ Yang-Mills to obtain  the non-perturbative corrections to the $\gamma$ profile.
The latter computed directly the corrections to the background by including
D(-1) branes in the configuration and resumming the string disc diagrams with any number of D(-1)-branes. They found a very compact expression for the (non-perturbatively) corrected profile of
$\gamma$,
\begin{equation}
\gamma=\gamma^{(0)}+\frac{i}{\pi}\left\langle\tr_{M}\log\frac{z-\ls^{-2}\varphi_1}{\mu}\right\rangle
-\frac{i}{\pi}\left\langle\tr_{M}\log\frac{z-\ls^{-2}\varphi_0}{\mu}\right\rangle\,,
\label{eqn:Lerdaetal}
\end{equation}
in terms of correlators of the quiver field theory. We will later arrive to this result in a simpler way without computing any string diagrams.
Moreover we will be able to evaluate explicitly these correlators.

Our goal is to to unravel the whole picture
that we have described hitherto from a purely microscopic description.

\section{The microscopic model}
\label{sec:micro}

In this section we detail the affine $A_1$ quiver theory governing the brane configuration on the $\CC^2/\ZZ_2$ orbifold,
paying special attention to the instanton sector that will be instrumental later on.
While the presentation we give makes use of string theory and D-branes, this is in no way necessary, as both the D3 and the D(-1)-branes' dynamics (in the ``near-horizon'' limit) can be described in field theory terms by gauge theories and instantons respectively.

\subsection{The four-dimensional gauge theory}

The field theory describing the dynamics (in the $\ls\to0$ field theory limit) of $N+M$ D3-branes of type~$0$ and $N+M$ D3-branes of type~$1$ is a four-dimensional superconformal $\nn=2$ gauge theory with gauge group $\U(N+M)_0\times\U(N+M)_1$ \cite{Lawrence:1998ja}.
Let us briefly recall how this field theory arises.

As is well known, $N+M$ D3-branes in flat space are described in field theory by the $U(N+M)$ $\nn=4$ supersymmetric Yang-Mills theory.
In terms of $\nn=1$ multiplets, the field content of this theory is one vector multiplet and three chiral multiplets $\Phi^I=(\Phi^1,\Phi^2,\Phi^3)$ transforming in the adjoint representation of the gauge group.
The superpotential of this model is given by
\begin{equation}
  W_{\nn=4}=\tr \Phi^1[\Phi^2,\Phi^3] \, ,
  \label{eqn:Wn4}
\end{equation}
using the same notation for a chiral superfield and the complex scalar which is its lowest component.
In the D-brane picture, these three complex scalars describe the fluctuations of the D3-branes in the six transverse directions, which can be paired to form the $\CC^3$ space.

We now want to replace this smooth transverse space by the $\CC\times \CC^2/\ZZ_2$ orbifold.
To this end, we let $\ZZ_2$ act on $\CC^3$ in the following way,
\begin{equation}
  g\cdot(z^1,z^2,z^3)=(z^1,-z^2,-z^3)
  \label{eqn:Z2action}
\end{equation}
where $g$ is the non-trivial element of $\ZZ_2$.
This yields the $\CC\times\CC^2/\ZZ_2$ orbifold by identifying all points of $\CC^3$ with their image under \eqref{eqn:Z2action}.
The first coordinate is fixed under the orbifold action, hence the orbifold singularity is a complex plane $\CC$ (times the Euclidean space-time $\RR^4$).
This plane will play an important role in the rest of the paper, and we will often write $z=z_1$ for conciseness.
To obtain the action of the orbifold on the fields, we also need to take into account the transformation of the endpoints of the open strings stretched between the D3-branes.
It turns out \cite{Lawrence:1998ja} that one needs to start with a gauge group whose rank is now twice the number of D3-branes (to take into account the orbifold images).
The element $g$ of $\ZZ_2$ is then represented on the Chan-Paton factors by the following matrix,
\begin{equation}
  R_{N+M}(g)=\begin{pmatrix}
    \I_{N+M} & 0 \\
    0 & -\I_{N+M}
  \end{pmatrix} \, .
  \label{eqn:RNg}
\end{equation}
This matrix corresponds to $N+M$ copies of the regular representation of $\ZZ_2$. 
Choosing this specific representation of $\ZZ_2$ corresponds to having (in the UV) $N+M$ regular D3-branes, i.e.\ a pair of images under $\ZZ_2$, which are then free to move in the full transverse space.
The orbifold action on the superfields $\Phi^I$ is then given by conjugation by $R_{N+M}$ combined with \eqref{eqn:Z2action},
\begin{equation}
  \Phi^I\to R_{N+M}(g) g\cdot\Phi^I R_{N+M}(g) \, .
  \label{eqn:phiorbifold}
\end{equation}
The modes that survive the orbifold projection are those which are invariant under \eqref{eqn:phiorbifold},
\begin{equation}
  \Phi^1= \begin{pmatrix}
    \varphi_0 & 0 \\
    0 & \varphi_1
  \end{pmatrix} \, , \quad
  \Phi^2= \begin{pmatrix}
    0 & \varphi^2_{01} \\
    \varphi^2_{10} & 0
  \end{pmatrix} \, , \quad
  \Phi^3= \begin{pmatrix}
    0 & \varphi^3_{01} \\
    \varphi^3_{10} & 0
  \end{pmatrix} \, .
  \label{eqn:phiproject}
\end{equation}
This projection also has the effect of breaking the gauge group $\U(2(N+M))\to\U(N+M)\times\U(N+M)$ as well as supersymmetry from $\nn=4$ to $\nn=2$.
In terms of $\U(N+M)_0\times\U(N+M)_1$ representations, one gets from \eqref{eqn:phiproject} one adjoint field for each $\U(N+M)$ factor, two bifundamental fields in the $((\mathbf{N+M})_0,(\mathbf{\overline{N+M}})_1)$ as well as two bifundamental fields in the complex conjugate representation.

Plugging \eqref{eqn:phiproject} back into the superpotential \eqref{eqn:Wn4} of $\nn=4$ yields the superpotential of the model we will work with,
\begin{equation}
  W_{\CC^2/\ZZ_2}=\tr \left[ \varphi_0 \left(\varphi^2_{01}\varphi^3_{10}-\varphi^3_{01}\varphi^2_{10}\right) + \varphi_1\left(\varphi^2_{10}\varphi^3_{01}-\varphi^3_{10}\varphi^2_{01}\right) \right] \, .
  \label{eqn:WZ2}
\end{equation}
The classical moduli space  of vacua $\mathcal M_{\text{cl}}$ is given by the critical points of the superpotential up to (complexified) gauge equivalence, $\mathcal M_{\text{cl}}=\{\d W=0\}/(\U(N+M)_0\times\U(N+M)_1)_{\CC}$.
We will be interested in the Coulomb branch, corresponding to giving \VEV s only to the adjoint fields $\varphi_0$ and $\varphi_1$.
Classically, one can choose a gauge in which the adjoint fields are diagonal.
A point on the Coulomb branch is then parameterized by the expectation value of their diagonal matrix elements,
\begin{equation}
  \langle\varphi_0\rangle = \ls^{-2}\diag(\tilde z_1,\ldots,\tilde z_{N+M}) \, , \quad \langle\varphi_1\rangle = \ls^{-2}\diag(z_1,\ldots,z_{N+M}) \, ,
  \label{eqn:vevclass}
\end{equation}
where we have written the matrix elements in terms of quantities having dimension of length in order to interpret them as D3-brane positions, and we need to identify two configurations differing by a permutation of eigenvalues.
The D-brane interpretation of this vacuum configuration is the following.
As we have discussed, the $\Phi^1$ coordinate corresponds to fluctuations of the D3-branes along the $\CC$ direction that is invariant under the orbifold action \eqref{eqn:Z2action}.
Once we orbifold, \eqref{eqn:phiproject} suggests that we now describe the positions of two different stacks of D3-branes along the $\CC$ direction.
This is indeed the case: $\varphi_0$ and $\varphi_1$ describe the fluctuations of fractional branes of type $0$ and $1$ respectively along the orbifold fixed locus.
Being free to choose \eqref{eqn:vevclass} arbitrarily then means that one can choose the position of the two types of fractional branes independently.
If the expectation values are completely generic, that is if no $z_i$ coincides with any $\tilde z_j$, the expectation values of the other fields have to vanish on $\mathcal M_{\text{cl}}$ and the fractional branes are stuck at the orbifold fixed locus.
When the positions of two branes of different type coincide, say $z_1=\tilde z_1$, one can see that a new branch of the classical moduli space opens up, the $(1,1)$ matrix element of the bifundamental fields are not required to vanish.
This corresponds to two fractional branes of different type forming a regular brane bound state.
This regular brane is then free to move away from the orbifold singularity.
Since we want a configuration of $N$ regular branes and $M$ fractional branes of each type, we need to have exactly $N$ pairs of eigenvalues of the two types coincide.
The vacuum that we want to consider, the enhançon vacuum, is further specified by the requirement of $\ZZ_M$ rotational symmetry and dependence on a single scale $|z_0|$,
\begin{equation}
  \langle\varphi_0\rangle=\ls^{-2}\diag(\underbrace{0,\ldots,0}_{\text{N times}},z_0\, \omega, z_0\, \omega^2,\ldots,z_0\, \omega^M) \, , \quad \langle\varphi_1\rangle=0 \, ,
  \label{eqn:vevenhanconvac}
\end{equation}
where $\omega$ is an $M$-th root of $-1$, $\omega^M=-1$.

At the quantum level, the $z_i$ and $\tilde z_j$ as defined by \eqref{eqn:vevclass} are not globally well-defined coordinates on the moduli space.
Instead, we need to use a set of independent gauge-invariant observables as coordinates of the Coulomb branch.
Rather than specifying those directly, we can instead encode the Coulomb branch vacuum in  a ratio of two polynomials of degree $M$, $T_r(z)=T_0(z)/T_1(z)$, where
\begin{equation}
  T_0(z)=\prod_{i=1}^{M+N}(z-\tilde z_i) \, , \qquad T_1(z)=\prod_{j=1}^{M+N}(z-z_j) \, .
  \label{T0T1def}
\end{equation}
Note that these $z_i$ and $\tilde z_j$ do \emph{not} coincide with the ones in \eqref{eqn:vevenhanconvac} in the quantum theory; they only agree perturbatively.
Instead, they are given by \VEV s of gauge invariant operators built from traces of $\varphi_0$ and $\varphi_1$.
From now on, we will assume that they are defined by \eqref{T0T1def} instead of \eqref{eqn:vevenhanconvac}.
Imposing the same constraints as on \eqref{eqn:vevenhanconvac} now requires \cite{Benini:2008ir}
\begin{equation}
  T_0(z)=z^N(z^M+z_0^M) \, , \qquad T_1(z)=z^{N+M} \, .
  \label{eqn:T0T1enhvac}
\end{equation}

\subsection{Adding a D(-1)-brane probe}

Our goal is to derive the full non-perturbative profile of the twisted supergravity $\gamma$ from field theory data.
By \eqref{eqn:holotaugs}, at the perturbative level, $\gamma$ is related to the gauge coupling of one of the two $\U(N+M)$ gauge groups.
However, it is not so clear how to extend this relation to the non-perturbative level, since one needs to choose a regularization scheme to define a coupling and it is not clear \emph{a priori} which scheme is appropriate for the holographic interpretation of $\gamma$ as a twisted supergravity field.
A way out of this problem is to relate instead $\gamma$ to an observable in the field theory.
This observable will turn out to be intimately related to the effective action for a probe fractional D(-1)-brane, which we are going to construct from a D(-1)/D3-brane system along the lines of \cite{Ferrari:2012nw}.

We enrich the set-up of D3-branes on the $\CC^2/\ZZ_2$ orbifold that we discussed previously by adding a single fractional D(-1)-brane of type 1.
The combined system is described by a partition function of the schematic form
\begin{equation}
  \mathcal Z = \int \d\mu_{\text{D3}} \d\mu_{\text{D(-1)}} e^{-S_{\text{D3}}-S_{\text{D(-1)}}} \, .
  \label{eqn:ZD3D-1}
\end{equation}
In addition to the functional integration over the D3-brane fields weighted by the four-dimensional gauge theory action $S_{\text{D3}}$ discussed above, there is now also an (ordinary) integral over the fractional D(-1)-brane moduli, with an action $S_{\text{D(-1)}}$ that we now detail.
This action describes the low-energy dynamics of the -1/-1 strings starting and ending on the D(-1)-brane and the -1/3 strings with one endpoint on the D(-1) and the other on a D3-brane as well as their couplings to the D3-brane fields.
The -1/-1 strings are uncharged under the $\U(N+M)_0\times\U(N+M)_1$ gauge group, whereas the 3/-1 strings with their endpoint on the D3-branes of either type transform in a fundamental representation of the corresponding $\U(N+M)$ gauge group and have charge $-1$ under the D(-1)-brane $\U(1)$ gauge group.

\begin{figure}
  \centering\usetikzlibrary{arrows,decorations.markings,positioning}
\pgfarrowsdeclaretriple{<<<}{>>>}{to}{to}
\pgfarrowsdeclaretriple{>>>}{<<<}{to reversed}{to reversed}
\begin{tikzpicture}
  [
  every path/.style={thick},
  group/.style={draw,fill=gray!30,minimum size=1cm},
  instanton/.style={rectangle,group},
  gauge/.style={circle,group}
  ]
  %Quiver nodes
  \node[instanton] (instanton 1) {$1$};
  \node[gauge] (gauge 1) [above=3 of instanton 1] {${N+M}$} node[above=\lineskip of gauge 1.north] {Type $1$};
  \node[gauge] (gauge 0) [left=4 of gauge 1] {${N+M}$} node[above=\lineskip of gauge 0.north] {Type $0$};
  %Adjoint fields
  \begin{scope}
    [decoration={markings,mark=at position 0.5 with {\arrow{<}}}]
    \draw [postaction={decorate}] (gauge 0.175) arc (5:355:0.5);
    \node [left=1 of gauge 0.west] {$\varphi_0$};
    \draw [postaction={decorate}] (instanton 1.east) arc (90:-180:0.5);
    \node [right=0.3 of instanton 1.east] {$\phi_1$};
    \draw [postaction={decorate}] (gauge 1.5) arc (175:-175:0.5);
    \node [right=1 of gauge 1.east] {$\varphi_1$};
  \end{scope}
  %Vertical arrows
  \begin{scope}
    [decoration={markings,mark=at position 0.3 with {\arrow{<<<}},mark=at position 0.9 with {\arrow{>>>}}}]
    \draw [postaction={decorate}] (instanton 1.north) -- (gauge 1.south) node [pos=0.25, right] {$(\tilde q^\alpha,\tilde \chi^0,\tilde \chi^1)$} node [pos=0.85, right] {$(q^\alpha,\chi^0,\chi^1)$};
  \end{scope}
  %Horizontal arrows
  \begin{scope}
    [decoration={markings,mark=at position 0.3 with {\arrow{<<}},mark=at position 0.8 with {\arrow{>>}}}]
    \draw [postaction={decorate}] (gauge 0.east) -- (gauge 1.west) node [pos=0.3, above] {$(\varphi^2_{10},\varphi^3_{10})$} node [pos=0.8, above] {$(\varphi^2_{01},\varphi^3_{01})$};
  \end{scope}
  %Diagonal arrows
  \begin{scope}
    [decoration={markings,mark=at position 0.2 with {\arrow{<<}},mark=at position 0.8 with {\arrow{>>}}}]
    \draw [postaction={decorate}] (instanton 1.north west) -- (gauge 0.south east) node [inner sep=15pt,pos=0.1,left] {$(\tilde\chi^2,\tilde\chi^3)$} node [inner sep=15pt,pos=0.7, left] {$(\chi^2,\chi^3)$};
  \end{scope}
\end{tikzpicture}
  \caption{The quiver of the $\CC^2/\ZZ_2$ orbifold with the UV brane configuration that we consider: $N+M$ D3-branes of each type corresponding to a $\U(N+M)_0\times\U(N+M)_1$ gauge group and 1 D(-1)-brane of type~$1$.}
  \label{fig:Z2fracquiver}
\end{figure}
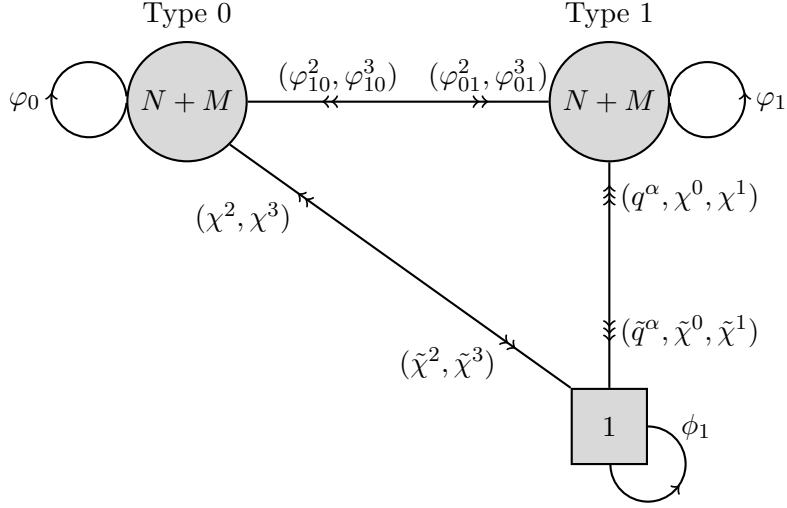

The action $S_{\text{D(-1)}}$ can be derived by a procedure similar to the one we followed for the four-dimensional gauge theory.
One starts with the action describing the D(-1)/D3 system in flat space \cite{Green:2000ke,Billo:2002hm} and one also embeds the $\ZZ_2$ orbifold group in the $\U(1)$ D(-1)-brane gauge group.
Since we are dealing with a fractional D(-1)-brane of type 1, the appropriate representation to take is not the regular representation as in \eqref{eqn:RNg}, but rather the non-trivial irreducible representation, $R^1(g)=-1 \in \U(1)$.
One then needs to truncate the moduli to the modes invariant under the orbifold action, which has a form similar to \eqref{eqn:phiorbifold} but with one (both) $R_{N+M}$ representation(s) replaced by $R^1$ for -1/3 strings (-1/-1 strings), as required by the $\U(N+M)_0\times\U(N+M)_1\times\U(1)$ representation to which they belong.
The various fields and moduli surviving the orbifold projection, and hence present in this brane configuration, are summarized in the quiver diagram of figure~\ref{fig:Z2fracquiver}.
The modulus $\phi_1$ is a complex number that plays the role of position of the fractional D(-1)-brane on the orbifold fixed plane $\CC$.
For dimensional reasons, we will rather work with the modulus $z$, related to $\phi_1$ by a rescaling,
\begin{equation}
  z=\ls^2\,\phi_1\,.
  \label{eqn:z}
\end{equation}
The -1/3 strings with their endpoints on the D3-branes of type 1 provide an $\SU(2)$ doublet of bosonic moduli $q^\alpha$ and their two fermionic superpartners $\chi^0$ and $\chi^1$, all in the anti-fundamental representation of $\U(N+M)_1$.
Similarly, the 3/-1 strings provide the complex conjugate moduli $(\tilde q^\alpha, \tilde\chi^0,\tilde\chi^1)$ in the fundamental of $\U(N+M)_1$.
On the other hand, the strings stretched between the D(-1)-brane and the D3-branes of type 0 only provide fermionic moduli, $(\chi^2,\chi^3)$ and $(\tilde\chi^2,\tilde\chi^3)$, which are charged under $\U(N+M)_0$.
The action $S_{\text{D(-1)}}$ reads
\begin{multline}
  S_{\text{D(-1)}} = -2\pi i \tau_1+ \frac{1}{4}\tr \left\{ (\tilde q^\alpha\phi^{\dagger}_1-\varphi^{\dagger}_1 \tilde q^\alpha)(\phi_1 q_\alpha -q_\alpha \varphi_1)  + (\tilde q^\alpha\phi_1-\varphi_1 \tilde q^\alpha)(\phi^{\dagger}_1 q_\alpha -q_\alpha \varphi^{\dagger}_1)  \right\} \\
  +\frac{1}{2}\tr \left\{ -(\tilde\chi^{1}_{}\phi^\dagger_{1}-\varphi^\dagger_{1}\tilde\chi^{1}_{})\chi^0 + \tilde\chi^0(\phi^\dagger_{1}\chi^1-\chi^1\varphi^\dagger_{1}) + \tilde\chi^3(\phi_1\chi^2 + \chi^2 \varphi_0) - \tilde\chi^2(\phi_1\chi^3 + \chi^3 \varphi_0) \right\} + \cdots \, .
  \label{eqn:SD-1mic}
\end{multline}
In this expression, we have already taken the ADHM/near-horizon limit \cite{Billo:2002hm,Ferrari:2012nw} and also dropped the terms involving the bifundamental fields since all their expectation values vanish on the Coulomb branch.
The fermionic neutral moduli $\psi$ (the superpartners of $\phi_1$) have not been written down explicitly either, since we are going to set them to $0$ anyway in the following.
The first constant term on the \RHS\ needs to be added in order to reproduce the instanton factor $e^{2\pi i\tau_1}$.

Let us comment on a subtlety concerning the gauge group of the theory in the near-horizon limit.
In this section, we have assumed it to be $\U(N+M)_0\times\U(N+M)_1\times\U(1)$, but the holographic dual describes only a $\SU(N+M)\times \SU(N+M)$ theory.
Indeed, the three commuting $\U(1)$ factors of the $\U(N+M)_0\times\U(N+M)_1\times\U(1)$ gauge group decouple.
The diagonal $\U(1)$ of the three gauge groups describes the movement of the full D(-1)/D3 system in the $\CC$-plane and, by translational invariance, no fields are charged under it.
The relative $\U(1)$ between the D(-1) and D3 gauge groups decouples since we will integrate out all the fields that are charged under it.
Finally, the anti-diagonal $\U(1)_B$ between the $\U(N+M)_0$ and $\U(N+M)_1$ groups is IR free and becomes a global baryonic symmetry.

The relevant global symmetry group of our field-theoretic model will then be $\SU(2)_R\times\SU(2)_F\times\U(1)_A\times\U(1)_B$. The last factor emerges from the near-horizon limit as we explained, and the other three can be seen to be the commuting subgroups of the parent $\SU(4)_R$ that survive the orbifolding. Geometrically, both $\SU(2)_R$ and $\SU(2)_F$ correspond to rotations in $\CC^2/\ZZ_2$ (the latter acts holomorphically on $\CC^2$, contrarily to the former), while the $\U(1)_A$ rotates the $\CC$ factor.

\section{The twisted supergravity field}
\label{sec:gamma}

Our system of background fractional D3-branes plus a probe D(-1)-brane is described by the
microscopic model we spelled out in the previous section. This model is governed by an action
of the form $S_{\text{D3}}+S_{\text{D(-1)}}$, where $S_{\text{D3}}$ is an $\nn=2$ action with superpotential \eqref{eqn:WZ2} and $S_{\text{D(-1)}}$ is written in \eqref{eqn:SD-1mic}.
We are interested in obtaining the effective action $S_{\text{D(-1),eff}}$ for the probe in the holographic background, in the
spirit of \cite{Ferrari:2012nw}:
\begin{equation}
\int \d\mu_{\text{D3}} \d\mu_{\text{D(-1)}}\,e^{-S_{\text{D3}}-S_{\text{D(-1)}}}=\int \d z \d \bar z \d \psi\,e^{-S_{\text{D(-1),eff}}} \, ,
\label{eqn:S.eff}
\end{equation}
where $z$ is the modulus defined in \eqref{eqn:z} and $\psi$ is its fermionic superpartner.
On the \RHS\ we are interested only in the bosonic part of the effective action, hence we can safely set $\psi=0$ as anticipated in \eqref{eqn:SD-1mic}.
The bosonic part of $S_{\text{D(-1),eff}}$ is equal to the twisted supergravity field $\gamma$, up to numerical factors. One way
to see this is by thinking of the fractional probe D(-1)-brane as a D1-brane wrapped on the
exceptional cycle $\Sigma$. For zero world-sheet gauge field, the Euclidean action of such an object is:
\begin{equation}
\frac{1}{\ls^2}\left(\int_{\Sigma}\mathrm{d}^2\xi\,e^{-\Phi}\sqrt{\det\left[P\left(G+B_2\right)\right]}-
i\int_{\Sigma} P\left(C_0\,B_2+C_2\right)\right)=-\frac{i}{\ls^2}\int_{\Sigma}\left(C_2+\left(C_0+i\,e^{-\Phi}\right)B_2\right)\,.
\end{equation}
Clearly, combining with the definition in \eqref{eqn:g.def}, we can write:
\begin{equation}
S_{\textrm{D(-1),eff}}=-2\pi i\,\gamma\,.
\label{eqn:S.g}
\end{equation}
We take the relation \eqref{eqn:S.g} as defining the twisted supergravity field outside the supergravity regime.
The first step to compute \eqref{eqn:S.eff} is to integrate out the fields that correspond to the degrees of freedom of the
D(-1)-D3 strings. In general this is done using large $N$ vector-model techniques (see 
\cite{Ferrari:2013pq,Ferrari:2013hg,Ferrari:2013wla} for examples and \cite{Ferrari:2013aba}
for a more general philosophy). In our case the integration can be done very simply since the action \eqref{eqn:SD-1mic} is quadratic in the moduli to be integrated out:
$q^{\alpha},\tilde{q}^{\alpha},\chi^0,\chi^1,\chi^2,\chi^3,\tilde{\chi}^0,\tilde{\chi}^1,
\tilde{\chi}^2,\tilde{\chi}^3$. Taking into account that the moduli with (without) a tilde are $(N+M)\times 1$ ($1\times(N+M)$) matrices, $\varphi_1$ and $\varphi_1^\dagger$ are adjoint fields with $(N+M)\times(N+M)$ components and $\phi,\phi^\dagger$ are $\CC$-number moduli, we can write the quadratic part of the action as 
\begin{equation}
S_{\text{D(-1)}}\supset\frac{1}{2}\left(q_{\alpha i}B\indices{^i_j}\tilde{q}^{\alpha j}+
\boldsymbol{\chi}_AF\indices{^A_B}\boldsymbol{\tilde{\chi}}^{B}\right)\,,
\end{equation}
where $\alpha=1,2$; $i,j$ go from $1$ to $N+M$ and $A,B$ go from $1$ to $4(N+M)$ because we have grouped the fermions as
\begin{equation}
  \boldsymbol{\chi}=\left(\chi^0_1,\ldots,\chi^0_{N+M},\chi^1_1,\ldots,\chi^2_1,\ldots,\chi^3_1,\ldots,\chi^3_{N+M}\right)\,,\qquad
  \boldsymbol{\tilde{\chi}}=\left(\tilde{\chi}^{0 1},\ldots,\tilde{\chi}^{3 N+M}\right)^{\operatorname T}\,.
\label{eqn:chi}
\end{equation}
The matrices $B$ and $F$ can be read from \eqref{eqn:SD-1mic}:
\begin{gather}
B\indices{^i_j}=\frac{1}{2}\left({\varphi_1}\indices{^i_k}{\varphi_1^\dagger}\indices{^k_j}+{\varphi_1^\dagger}\indices{^i_k}{\varphi_1}\indices{^k_j}\right)+\phi_1^\dagger\phi_1\delta\indices{^i_j}-\phi_1^\dagger{\varphi_1}\indices{^i_j}-{\varphi^\dagger}\indices{^i_j}\phi_1\,,\\
F=\left(\begin{array}{cccc}
0 & \phi_1^{\dagger}\mathbf{1}-\varphi_1^{\dagger} & 0 & 0\\
-\phi_1^{\dagger}\mathbf{1}+\varphi_1^{\dagger} & 0 & 0 & 0\\
0 & 0 & 0 & -\phi_1\mathbf{1}+\varphi_0 \\
0 & 0 & \phi_1\mathbf{1}-\varphi_0 & 0
\end{array}\right)\,,
\end{gather}
where we have written $F$ in $(N+M)\times(N+M)$ blocks and $\mathbf{1}$ represents the identity in each of these blocks. We notice that we can represent $B$ in matrix form as
\begin{equation}
B=\left(\phi_1\mathbf{1}-\varphi_1\right)\left(\phi_1^{\dagger}\mathbf{1}-\varphi_1^{\dagger}\right)+[\varphi_1,\varphi_1^\dagger]\,.
\label{eqn:B}
\end{equation}
As we did for the bifundamental fields, we can drop the last term $[\varphi_1,\varphi_1^\dagger]$ in \eqref{eqn:B} because it vanishes inside all correlators by the D-flatness condition for the gauge group $\SU(N+M)_1$ (with bifundamentals set to zero).
The result of the integration of the $q$ and $\chi$ moduli will be given by the ratio $\det F/\det B^2$ of the determinants of $F$ and $B$ (squared, because of the two indices of $q_\alpha$ and $\tilde q^\alpha$), which we can readily compute:
\begin{align}
\det\left(B\right)^2&=\det\left(\phi_1\mathbf{1}-\varphi_1\right)^2\det\left(\phi_1^{\dagger}\mathbf{1}-\varphi_1^{\dagger}\right)^2\,,\label{eqn:referee}\\
\det\left(F\right)&=\det\left(\phi_1\mathbf{1}-\varphi_0\right)^2\det\left(\phi_1^{\dagger}\mathbf{1}-\varphi_1^{\dagger}\right)^2\,.
\end{align}
Notice that when taking the quotient the dependence of the resulting expression on the D3 fields will be holomorphic, because the factors with $\phi_1^{\dagger}\mathbf{1}-\varphi_1^{\dagger}$ cancel between the bosonic and fermionic determinants. 
This will be key to performing the functional integral over the four-dimensional degrees of freedom. From
\eqref{eqn:S.eff}, taking into account also the first term in \eqref{eqn:SD-1mic}, we can write the integral to be performed as:
\begin{equation}
\int\d\mu_{\text{D3}} \d\mu_{\text{D(-1)}}\,\frac{\det\left(\phi_1\mathbf{1}-\varphi_0\right)^2}{\det\left(\phi_1\mathbf{1}-\varphi_1\right)^2}
\,e^{2\pi i\tau_1-S_{\text{D3}}}=\int \d\phi_1\d\bar\phi_1\d\psi\,e^{2\pi i\tau_1}
\left\langle\frac{\det\left(\phi_1\mathbf{1}-\varphi_0\right)^2}{\det\left(\phi_1\mathbf{1}-\varphi_1\right)^2}\right\rangle_{D3}\,.
\label{eqn:S.eff.2}
\end{equation}
If we use the identification \eqref{eqn:z} and we also rescale the fields $\varphi_i\to Z_i=\ls^2\varphi_i$,
we can write the following expression for $\gamma$ in terms of correlators in the quiver gauge theory:
\begin{equation}
e^{2\pi i\,\gamma}=e^{2\pi i\tau_1}\left\langle\frac{\det\left(z-Z_0\right)^2}{\det\left(z-Z_1\right)^2}\right\rangle=
e^{2\pi i\tau_1}\frac{\left\langle\det\left(z-Z_0\right)\right\rangle^2}{\left\langle\det\left(z-Z_1\right)\right\rangle^2}\,,
\label{eqn:g.corr}
\end{equation}
where we have used chiral factorization in the second equality, and we drop the explict $\mathbf{1}$ from now on.
This is a beautiful formula illustrating the emergence phenomenon, showing how the profile of the 
twisted supergravity field emerges from a ``microscopic'' quantity. What is even more striking is the
fact that we can compute these chiral correlators exactly. This is possible thanks to the recent remarkable
works \cite{Nekrasov:2012xe,Fucito:2012xc}, that extended the Seiberg-Witten technology to $\nn=2$
quivers.

Before writing the exact expression for the correlators, let us make contact with the formula
\eqref{eqn:Lerdaetal} obtained non-perturbatively  by string theory techniques \cite{Billo:2012st}. The computation of the $(N+M)\times (N+M)$ determinants of operators in \eqref{eqn:g.corr} entails a regularization scheme.
The natural way to define them is via the Fredholm determinant:
\begin{equation}
\det\left(z-Z\right)=\exp\left[\tr\log\left(z-Z\right)\right]\,.
\label{eqn:Fredholm}
\end{equation}
In this formula, both the exponential and the logarithm are to be understood as defined by their Taylor
series. When we act with a \VEV\ on the \LHS\ of \eqref{eqn:Fredholm}, because of chiral factorization,
on the \RHS\ we can act with the \VEV\ directly in the argument of the exponential.
So more explicitly for the case that concerns us, we write 
\begin{equation}
\left\langle\det\left(z-Z_a\right)\right\rangle^2=\exp\left[2\left\langle\tr\log\left(z-Z_a\right)\right\rangle\right]\,.
\label{eqn:Fredholm.VEV}
\end{equation}
Using \eqref{eqn:Fredholm.VEV} in \eqref{eqn:g.corr}, plus the fact that $\gamma^{(0)}$ equals the bare coupling $\tau_1$, we easily recover \eqref{eqn:Lerdaetal}.

This was to be expected, but the reader might be befuddled by the following puzzling aspect: the
computation of the twisted supergravity profile in \cite{Billo:2012st}, leading to \eqref{eqn:Lerdaetal},
involves resumming a series of string amplitudes encoding the interaction among D3 and D(-1) branes.
We are instead performing a simple Gaussian integration to arrive at the result.
The authors of \cite{Billo:2012st} essentially follow the opposite approach to ours.
They want to obtain non-perturbative corrections to the $\gamma$-profile \eqref{eqn:g.pert} by adding $k$ fractional D(-1)-branes to the D3-brane set-up (yielding a $\U(k)$ non-Abelian generalization of \eqref{eqn:SD-1mic}) and integrating them out.
On the one hand, these branes couple to $\gamma$. 
On the other hand, they can be interpreted as gauge theory instantons, relating in this way instanton corrections in gauge theory to corrections to the $\gamma$-profile.
Resumming the contributions for all values of $k$ then yields \eqref{eqn:Lerdaetal}.
On the contrary, our approach is to keep the D(-1)-brane and integrate out the D3-branes, yielding immediately the full gauge-theory correlator in \eqref{eqn:S.eff.2} with no need to make explicit nor resum the instanton series that contributes to it.

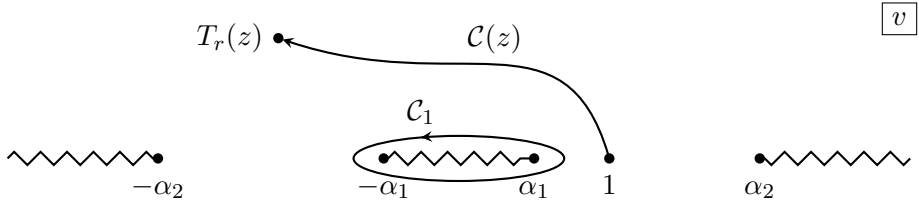
\begin{figure}
    \centering\usetikzlibrary{arrows,decorations.pathmorphing,positioning}
\begin{tikzpicture}
  [
  scale=2,
  every path/.style={thick},
  cut/.style={draw,decoration=zigzag, decorate},
  point/.style={shape=circle,draw,fill,scale=0.3},
  contour/.style={draw,->,>=stealth},
  ]
  \node (corner) at (3,1){} node[draw,thin] at (corner.south west) {$v$};
  %branch points
  \node[point] (-a2) at (-2,0) {} node [below=0.1 of -a2] {$\smash{-}\alpha_2$};
  \node[point] (a2) at (2,0) {} node [below=0.1 of a2] {$\alpha_2$};
  \node[point] (-a1) at (-0.5,0) {} node [below=0.1 of -a1] {$\smash{-}\alpha_1$};
  \node[point] (a1) at (0.5,0) {} node [below=0.1 of a1] {$\alpha_1$};
  % other points
  \node[point] (one) at (1,0) {} node [below=0.1 of one] {$\smash{1}\vphantom{\alpha_1}$}; %dirty alignment trick
  \node[point] (z) at (-1.2,0.8) {} node [left=0 of z.west] {$T_r(z)$};
  %branch cuts
  \path[cut] (-3,0) -- (-a2);
  \path[cut] (3,0) -- (a2);
  \path[cut] (-a1) -- (a1);
  %paths
  \path[contour] (one) .. controls (0.7,1) and (0,0.4) .. (z) node [above,midway] {$\mathcal C(z)$};
  \draw (0,0) ellipse (0.7 and 0.15);
  \path[contour] (110:0.7 and 0.15) arc(110:111:0.7 and 0.15) node [above=\lineskip] {$\mathcal C_1$};
\end{tikzpicture}
    \caption{
       The data specifying the integral \eqref{eqn:gammaintv} on the Riemann sphere $\CC\cup\{\infty\}$ with coordinate $v$.
       The integrand has two branch cuts: between $-\alpha_1$ and $\alpha_1$ and between $-\alpha_2$ and $\alpha_2$ through $\infty$.
       The ambiguity in the choice of $\mathcal C(z)$ is characterized by the integral \eqref{eqn:gammaCCp} along $\mathcal C_1$.
     }
    \label{fig:intgamma}
\end{figure}

Let us state now the final expression for the correlator in \eqref{eqn:g.corr}, leaving
all the details on how to extract it from \cite{Nekrasov:2012xe,Fucito:2012xc} for the appendix.
As usual when one deals with instantons, the result is more conveniently expressed in terms of the
variables:
\begin{equation}
\qq_a= e^{2\pi i\,\tau_a}=e^{-\frac{8\pi^2}{g_a^2}}\,e^{i\,\vartheta_a}\,,\qquad \qq=\qq_0\,\qq_1\,.
\label{eqn:qs}
\end{equation}
A contribution from a $k$-instanton of type $a$ comes with a factor $\qq_a^k$. Recalling
from the previous section that a point on the Coulomb branch of the quiver theory is specified by the
quotient of two monic polynomials $T_r=T_0/T_1$; in such a vacuum our correlator turns out to be:
\begin{equation}
2\pi i\left(\gamma-\gamma^{(0)}\right)=\beta(\qq_a)\int_z^{\infty}\mathrm{d}x
\,\frac{T'_r(x)}
{\sqrt{T_r(x)^2-\alpha_1(\qq_a)^2}\sqrt{T_r(x)^2-\alpha_2(\qq_a)^2}}\,,
\label{eqn:corr}
\end{equation}
where $\gamma^{(0)}=\tau_1$ and the precise definitions of $\alpha_1,\alpha_2$ and $\beta$ can be found in the appendix. For the discussion
that follows, it is enough to know that these quantities are well-behaved functions admitting a small
$\qq_a$ expansion:
\begin{equation}
\begin{aligned}
\beta(\qq_a)&=-\frac{i}{\sqrt{\qq_1}}\left(1+\qq_1-\qq_0+6\qq+\qq_0^2+\OO\left(\qq_a^3\right)\right)\,,\\
\alpha_1(\qq_a)&=2\sqrt{\qq_0}\left(1+\qq_1-\qq_0-6\qq+\qq_0^2+\OO\left(\qq_a^3\right)\right)\,,\\
\alpha_2(\qq_a)&=\frac{1}{2\sqrt{\qq_1}}\left(1+\qq_1-\qq_0+10\qq+\qq_0^2+\OO\left(\qq_a^3\right)\right)\,.
\end{aligned}
\label{eqn:baa}
\end{equation}
It is generally more convenient to change integration variables in \eqref{eqn:corr} to $v=T_r(x)$,
\begin{equation}
2\pi i\left(\gamma-\gamma^{(0)}\right)=-\beta\,\int_1^{T_r(z)}
\frac{\mathrm{d}v}{\sqrt{(v^2-\alpha_1^2)(v^2-\alpha_2^2)}}\,.
\label{eqn:gammaintv}
\end{equation}
The contour of integration in \eqref{eqn:gammaintv} and the branch cut structure of the integrand are represented in figure~\ref{fig:intgamma}.
The contour of integration $\mathcal C(z)$ must go from $v=1$ (corresponding to $z=\infty$) to $v=T_r(z)$ without crossing any branch cuts, but is otherwise arbitrary.
This does not fix $\gamma(z)$ unambiguously: one can choose a contour encircling the branch cut between $-\alpha_1$ and $\alpha_1$ an arbitrary number of times and the value of the integral will depend on this number.
If the contour $\mathcal C'(z)$ makes one more counter-clockwise circle around the cut than $\mathcal C(z)$, the difference in the resulting $\gamma$ functions is
\begin{equation}
  2\pi i (\gamma_{\mathcal C'}-\gamma_{\mathcal C})=-\beta\,\int_{\mathcal C_1}
\frac{\mathrm{d}v}{\sqrt{(v^2-\alpha_1^2)(v^2-\alpha_2^2)}} = 4\pi i\, .
  \label{eqn:gammaCCp}
\end{equation}
This has no physical significance as is obvious from \eqref{eqn:g.corr}. From
\eqref{eqn:g.def} we can also see that such a shift corresponds to a shift of $c$, or equivalently
a $4\pi$-shift of the $\vartheta$-angle in the field theory, which is of course not observable.
Similarly, the path can encircle the other branch cut an arbitrary number of times.
Since $\mathcal C_1$ can be continuously deformed into a path going around the branch cut between $\alpha_2$ and $-\alpha_2$ clockwise, this also corresponds to a non-observable $4\pi$ shift in the $\vartheta$-angle.

As a quick check of the formula \eqref{eqn:gammaintv}, we can recover the perturbative result \eqref{eqn:g.pert}.
What we have to do is to send $\qq_a\to0$, keeping only the leading order. Then
\begin{equation}
\beta\to-\frac{i}{\sqrt{\qq_1}}\,,\qquad\sqrt{v^2-\alpha_1^2}\to v\,,\qquad\sqrt{v^2-\alpha_2^2}\to\frac{i}{2\sqrt{\qq_1}}\,,
\end{equation}
which gives the trivial integral:
\begin{equation}
\pi i\left(\gamma-\gamma^{(0)}\right)=\int_1^{T_r(z)}\frac{\mathrm{d}v}{v}=
\log\left(T_r(z)\right)\, .
\end{equation}
When we use \eqref{T0T1def}, this is precisely the perturbative formula \eqref{eqn:g.pert} we were expecting. 

A less trivial check is to send only $\qq_0\to0$, but keep $\qq_1$ arbitrary.
This corresponds to suppressing the dynamics of the type~0 gauge group, which in this limit plays the role of a global flavor group.
Hence the theory one obtains is $\nn=2$ SQCD with $2M$ flavors\footnote{Recall that $\gamma$ is completely insensitive to the $N$ regular branes.} on the Coulomb branch.
This is exactly the regime considered in \cite{Billo:2012st,Martucci:2012jk} and we can compare our formula \eqref{eqn:gammaintv} for $\gamma$ in this limit with theirs.
Setting $\qq_0=0$, the expansions \eqref{eqn:baa} truncate to
\begin{equation}
  \beta(0,\qq_1)=-\frac{i}{\sqrt{\qq_1}}(1+\qq_1) \, , \quad \alpha_1(0,\qq_1)=0, \, \quad \alpha_2(0,\qq_1)=\frac{1}{2\sqrt{\qq_1}}(1+\qq_1) \, .
  \label{eqn:baa.SQCD}
\end{equation}
The integral in \eqref{eqn:gammaintv} then reduces to
\begin{equation}
  2\pi i\left(\gamma-\gamma^{(0)}\right)=-\beta(0,\qq_1)\int_1^{T_r(z)}
  \frac{\mathrm{d}v}{v\sqrt{v^2-\alpha_2(\qq_1,0)^2}}
  =\frac{i\beta}{2\alpha_2}\left. \log \left( \frac{1-\sqrt{1-v^2/\alpha_2^{2}}}{1+\sqrt{1-v^2/\alpha_2^{2}}} \right) \right|_1^{T_r(z)}\, .
  \label{eqn:gammaSQCDcomp}
\end{equation}
Using \eqref{eqn:baa.SQCD} and some elementary algebra, the lower bound contribution is found to be $-\log \qq_1=-2\pi i \gamma^{(0)}$, hence
\begin{equation}
  2\pi i\,\gamma(z)
  = \log \left( \frac{1-\sqrt{1-T_r(z)^2/\alpha_2^{2}}}{1+\sqrt{1-T_r(z)^2/\alpha_2^{2}}} \right) \, ,
  \label{eqn:gammaSQCD}
\end{equation}
in perfect agreement with the result of \cite{Billo:2012st,Martucci:2012jk}.

\section{Large $N$ limit. The enhançon}
\label{sec:enhancon}

The expression \eqref{eqn:gammaintv} we wrote for the twisted supergravity field $\gamma$ (taking \eqref{eqn:g.corr} as its definition) is completely general, since
it has been derived from the field theory in full non-perturbative glory. In particular, it is valid all along the
RG flow for any point on the Coulomb branch, for any value of the couplings and any integer numbers $N,M$.
Looking at it from the string theory
perspective, it means that \eqref{eqn:gammaintv} contains all $g_s$ and $\alpha'$ corrections to the dynamics
of the brane array we are considering. However, for the time being we are only interested in using a small fraction
of this power. We consider small $g_s$ and large $N,M$, corresponding to the supergravity regime. For
convenience, we assume that the two bare gauge couplings are equal and that $N$ is proportional to $M$:
\begin{equation}
  \qq_0=\qq_1=e^{-\frac{\pi}{g_s}} \, , \quad N=p\,M\,,\quad p\in\mathbb{Q}\,.
\label{eqn:NM}
\end{equation}
As we discussed in Section \ref{ssec:enh}, in this regime a curious phenomenon is taking place, that of the
enhançon. While the field-theoretical mechanism behind it is understood (recall it has to do with the impossibility
of bringing the roots of the Seiberg-Witten curve to the origin of the moduli space) and its effect in the
supergravity background (the need for an excision procedure below a certain scale) is also well-known, as far as we know
there is no fully general construction in the literature explaining the interplay of these aspects. We hope to
fill this gap here. The idea is to solve the integral \eqref{eqn:gammaintv} for different vacua, characterized by different functions
$T_r$, and analyze their large $M$ limit.

The large $M$ limit corresponds to taking $M\to\infty$ and $\qq_a\to0$, keeping the 't Hooft couplings $\lambda_a$ defined by \eqref{eqn:holo1} fixed, which translates to keeping
\begin{equation}
  \qq_a^{\frac{1}{M}}=e^{-\frac{8\pi^2(p+1)}{\lambda_a}}=e^{-\frac{\pi}{g_s M}} \quad\text{fixed} \, .
  \label{eqn:qalargeM}
\end{equation}
If we furthermore wanted to suppress the $\alpha'$ corrections and obtain two-derivative gravity we should take the limit $\lambda_a\to\infty$ in which \eqref{eqn:qalargeM} goes to~$1$.
We will however refrain from taking this limit, as it eliminates the separation between the scale $|z_0|$ at which the theory is Higgsed and the enhançon scale which, as we will see, is $\sim\qq_1^{l/M}|z_0|$ for some finite number $l$ that does not scale with $M$.

In the large large $M$ limit, $\alpha_1\to0$ by \eqref{eqn:baa}, and it seems that we can replace in \eqref{eqn:gammaintv} $v^2-\alpha_1^2$ by $v^2$.
This is not always true, depending on the value of the upper bound $T_r(z)$.
If $T_r(z)$ stays at a finite distance from $\pm\alpha_1$ in the large $M$ limit, one can choose an integration contour $\mathcal C(z)$ as in figure~\ref{fig:intgamma} that stays away from the branch points $\pm\alpha_1$ and this approximation is valid.
The computation of $\gamma$ then reduces to \eqref{eqn:gammaSQCD}, where one now has to take the large $M$ limit.
In other words, the large $M$ limit of this model reduces generically (in the sense we just discussed) to the large $M$ limit of $\nn=2$ SQCD with $2M$ flavors, a result which was already anticipated by \cite{Benini:2008ir} from the study of the Seiberg-Witten curve, but that we have now shown directly on the twisted supergravity field.
Whether this condition on $T_r(z)$ is satisfied depends both on the Coulomb branch vacuum encoded by $T_r$ and the specific $z$ considered.
If it fails, one needs to do a more refined analysis, similarly to what was done in \cite{Ferrari:2001mg} for pure $\nn=2$ Yang-Mills theory.
For a given vacuum $T_r$, we will call the points that satisfy the condition ``\emph{ordinary points}'' and ``\emph{exceptional points}'' the ones that do not.

\subsection{The enhançon vacuum}

\begin{figure}[htb]
  \centering\usetikzlibrary{arrows,decorations.pathmorphing,decorations.markings,positioning,spy}
\newcommand*{\qscale}{0.1} %2\sqrt{\qq}
\newcommand*{\numM}{10} %M=10
% \pgfmathsetmacro{\scaleaip}{1/(1+\qscale)^(1.0/\numM)} %|1+\alpha_1|^{-1/M}
\pgfmathsetmacro{\scaleaip}{0.990514} %|1+\alpha_1|^{-1/M}
% \pgfmathsetmacro{\scaleaim}{1/(1-\qscale)^(1.0/\numM)} %|1-\alpha_1|^{-1/M}
\pgfmathsetmacro{\scaleaim}{1.01059} %|1-\alpha_1|^{-1/M}
% \pgfmathsetmacro{\scaleaiip}{1/(1/\qscale+1)^(1.0/\numM)} %|1+\alpha_2|^{-1/M}
\pgfmathsetmacro{\scaleaiip}{0.786793} %|1+\alpha_2|^{-1/M}
% \pgfmathsetmacro{\scaleaiim}{1/(1/\qscale-1)^(1.0/\numM)} %|1-\alpha_2|^{-1/M}
\pgfmathsetmacro{\scaleaiim}{0.802742} %|1-\alpha_2|^{-1/M}
\begin{tikzpicture}
  [
  scale=4,
  every path/.style={thick},
  point/.style={fill,circle,scale=0.15},
  coord/.style={thin,black,fill=white, inner sep=0},
  spy using outlines={circle, magnification=4, size=5cm, connect spies, every spy in node/.style={draw,fill=white}},
  ]
  \draw[coord] circle(1);
  \draw[coord] circle(0.794328);
  \node[coord,below=0.2] {$0$};
  \node[point] at (0,0) {};
  \node [draw,thin,below left,rectangle] at (2.3,1) {$z$};
  \foreach \x in {0,...,9} 
  {
    \draw[black] (18+36*\x:1) node[point] {}; 
    \draw[red] (18+36*\x:\scaleaip) node[point] {} -- (18+36*\x:\scaleaim) node[point] {};
    \draw[blue,decoration={zigzag,segment length=0.22cm,amplitude=2pt},decorate] (36*\x:\scaleaiip) node[point] {} -- (18+36*\x:\scaleaiim) node[point] {};
  }
 \draw[->,>=stealth] (2.5,0) -- (0.794328+0.05,0) arc (0:-180:0.05) -- (0.5,0) node[point] {} node[below] {$z$} ;
  \node[coord,xshift=5pt] at (170:0.794328) {$|z_0|{\mathfrak q}_1^{\frac{1}{2M}}$};
  \node[coord] at (170:1) {$|z_0|$};
  \spy[black!70!green] on (18:0.9*4) in node [below right] at (1.4,0.2);
\end{tikzpicture}
  \caption{The singularity structure in the $z$ plane of the integrand in \eqref{eqn:int.enh} for the enhançon vacuum with $M=10$, $\sqrt{\qq_1}=0.05$.
  Part of the figure has been enlarged for better visibility. The black dots are the roots of $T_0$ and $T_1$ corresponding to the classical positions of the fractional D3-branes. The red (blue) dots are the branch points at the scale $\sim|z_0|$ (at the enhançon radius) in the first (second) column of \eqref{eqn:branchpoints} and the very short red (zigzag blue) lines are the branch cuts joining them.
The integration path going from $\infty$ to $z$ is real except near the branch point at the enhançon radius.}
  \label{fig:enhancon}
\end{figure}
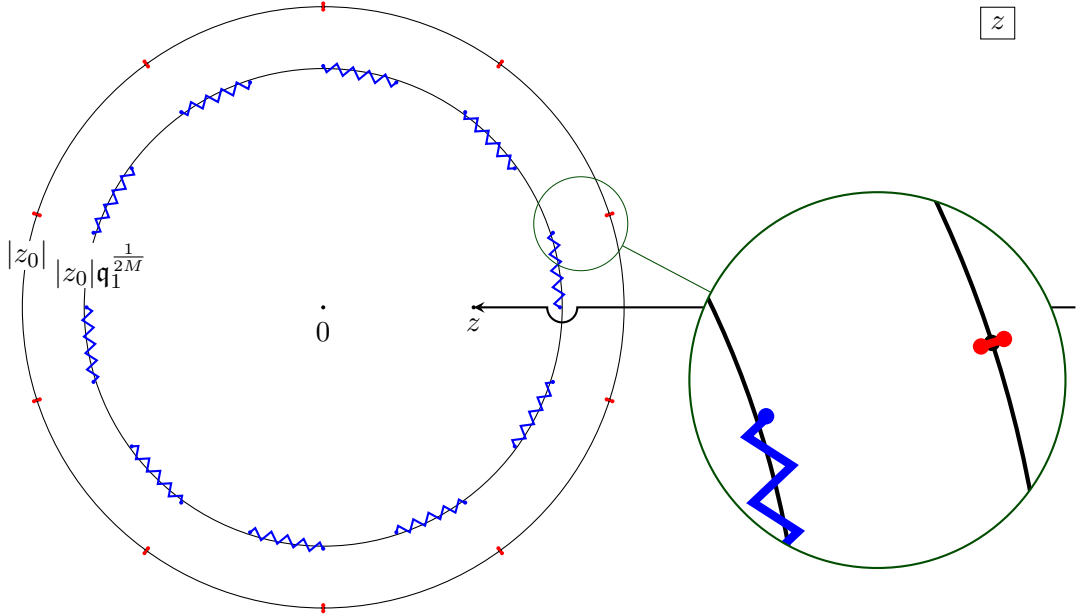

Let us first focus on arguably the simplest brane array: the one that corresponds classically to $M$ fractional
branes of type 0 distributed on a circle of radius $|z_0|$ and $M$ fractional branes of type 1 at the origin, where the
$N$ regular branes sit too. Of course, as we have already mentioned, this picture is corrected non-perturbatively,
where anyway it does not make sense to talk about brane positions. The way we characterize the configuration is by:
\begin{equation}
T_0=z^N\left(z^M+z_0^M\right)\,,\qquad T_1=z^{N+M}\,\quad\implies\quad T_r=1+\left(\frac{z_0}{z}\right)^M\,.
\label{eqn:enh.vac}
\end{equation}
Plugging this $T_r$ into the formula for $\gamma$ \eqref{eqn:gammaintv}, and using the rescaled variable $u=\frac{z_0}{z}$,
we can write
\begin{equation}
2\pi i\left(\gamma-\gamma^{(0)}\right)=-\beta\int_0^{\frac{z_0}{z}}\mathrm{d}u\,
\frac{M\,u^{M-1}}{\sqrt{(u^M+1)^2-\alpha_1^2}\,\sqrt{(u^M+1)^2-\alpha_2^2}}\,,
\label{eqn:int.enh}
\end{equation}
where we should recall that $\beta$, $\alpha_1$ and $\alpha_2$ depend on $\qq_a$. The contour of integration
must be chosen so as not to cross any branch cuts. Let us take for definiteness $z_0,z,\qq_a\in\RR$. Recall
that we want to work with small $g_s$, or equivalently small $\qq_a$, subject to the condition \eqref{eqn:qalargeM}. Given the
expansions in \eqref{eqn:baa}, we see that the branch points are located at
\begin{equation}
\begin{aligned}
&z=z_0\left|1+\alpha_1\right|^{-1/M}\omega^{k+\frac{1}{2}}\,, &&z=z_0\left|\alpha_2+1\right|^{-1/M}\omega^k\,,&&k=0\,\ldots M-1\,,\\
&z=z_0\left|1-\alpha_1\right|^{-1/M}\omega^{k+\frac{1}{2}}\,, &&z=z_0\left|\alpha_2-1\right|^{-1/M}\omega^{k+\frac{1}{2}}\,,&&k=0\,\ldots M-1\,,
\end{aligned}
\label{eqn:branchpoints}
\end{equation}
with $\omega=e^{\frac{2\pi i}{M}}$ an $M$-th root of unity. We take the branch cuts to link the branch points
sharing a column in \eqref{eqn:branchpoints}. These branch cuts are represented in figure~\ref{fig:enhancon}.
The first set of $2M$ branch
points are very close to the classical roots $z_0\omega^{k+\frac12}$. The length of these branch cuts is approximately
$\frac{2z_0\alpha_1}{M}$, which, using the identifications \eqref{eqn:holo1} and \eqref{eqn:qs}, is seen to be of order
$\OO\left(\frac{1}{M}\,e^{-\frac{4\pi^2(p+1)}{\lambda}M}\right)$. Such exponentially small branch cuts in the large $M$ limit can be associated with sharply localized D-branes. 
This matches precisely the expectations coming from supergravity, that is, to find $M$ D3-branes of type 0 at
the positions $z_0\,w^{k+\frac12}$. The second set of $2M$ branch points is quite different. They fill homogeneously a pair of circles of radius
\begin{equation}
|z_0|\left(\frac{1}{2\sqrt{\qq_1}}\pm1\right)^{-\frac{1}{M}}\approx|z_0|\,\qq_1^{\frac{1}{2M}}=|z_0|\,e^{-\frac{\pi}{2g_sM}}\,,
\label{eqn:enh.FT}
\end{equation}
and the distance between consecutive branch points is of order $\OO\left(\frac{1}{M}\right)$. Notice that the scale above
is exactly the enhançon scale \eqref{eqn:rho_1} arising from supergravity considerations. We will comment more on
this below. The length of the branch cuts is now of the same order as the separation between the cuts and the concept
of D3-brane is lost at this scale.

In view of \eqref{eqn:branchpoints}, we see that we can take the contour with $u$ real in \eqref{eqn:int.enh}. This
will not cross any branch cut. For small enough $z$, this path hits the branch point at $|z_0|\,e^{-\frac{\pi}{2g_sM}}$
but we can always go just below the cut as in figure~\ref{fig:enhancon}.

Let us now study the large $M$ limit of the integral \eqref{eqn:int.enh}.
The first step is to identify the exceptional points. By definition, they coincide with the branch points in the first column of \eqref{eqn:branchpoints} in the large $M$ limit and by consequence, the formula \eqref{eqn:gammaSQCD} does not apply to them.
Since in the large $M$ limit, the branch points densely fill the ring of radius $|z_0|$, the points with $|z|=|z_0|$ are exceptional.
All the points with fixed $z, |z|\neq|z_0|$ are then ordinary.
A more general way to construct an exceptional point is to scale its coordinate in the large $M$ limit, taking $z=z_b+w/M$ with $z_b$ a branch point in the first column of \eqref{eqn:branchpoints} and $w$ fixed in the large $M$ limit.
We will however not pursue this possibility here, since we are interested in the profile of $\gamma$ as a function of $z$ and the enhançon mechanism which happens at a scale \eqref{eqn:enh.FT}, well separated from $|z_0|$ at finite 't~Hooft coupling.

For ordinary points, the formula \eqref{eqn:gammaSQCD} is valid and we have
\begin{equation}
  2\pi i\, \gamma(z)=
  \log \left( \frac{1-\sqrt{1-T_r(z)^2/\alpha_2^{2}}}{1+\sqrt{1-T_r(z)^2/\alpha_2^{2}}} \right)\, , \quad T_r(z)=1+ \left( \frac{z_0}{z} \right)^M \, , \quad \frac{1}{\alpha_2^2}=4\qq_1 \, .
  \label{eqn:easy}
\end{equation}
Since $\alpha_2$ is very large, we can simplify further the expression for $\gamma$. But
for that, we have to be careful with the range of $z$. Let us distinguish three regions:

\paragraph{Region ${|z|>|z_0|}$ :}
This region corresponds to the UV of the field theory, where we have a gauge group $\SU(N+M)\times\SU(N+M)$
and we expect conformality. Therefore $\gamma$ should not run. Here $|T_r(z)|\ll|\alpha_2|$, and we can approximate:
\begin{equation}
\sqrt{1-T_r(z)^2/\alpha_2^2}\approx 1 - \frac{T_r(z)^2}{2\alpha_2^2} \,.
\label{eqn:approx}
\end{equation}
Plugging this into \eqref{eqn:easy}, we obtain for the twisted field
\begin{equation}
\gamma=\gamma^{(0)}-\frac{i}{\pi}\log\left[1+\left(\frac{z_0}{z}\right)^M\right]\,.
\label{eqn:g.p.enh}
\end{equation}
For large $M$, we can neglect the second term inside the logarithm, and we obtain the desired result:
\begin{equation}
\gamma=\gamma^{(0)}=\frac{i}{2g_s}\,.
\end{equation}

\paragraph{Region ${|z_0|>|z|>\rho_1}$ :}
This is the region where the approximation $|T_r(z)|\ll|\alpha_2|$ is still valid. Recall from \eqref{eqn:enh.FT}
that its boundary is at the scale $\rho_1$ of \eqref{eqn:rho_1}, where supergravity was predicting
the enhançon phenomenon. Given that the approximation is the same one used above, we can follow the
reasoning there, to arrive to the expression \eqref{eqn:g.p.enh}. In this case though, the term we 
have to neglect inside the logarithm is the first one. This gives
\begin{equation}
\gamma=\gamma^{(0)}+\frac{i\,M}{\pi}\log\frac{z}{z_0}\,.
\end{equation}
Again no surprises here. This is the expected result \eqref{eqn:g.enh}.

\paragraph{Region ${|z|<\rho_1}$ :}
This is the region where the supergravity background ceases to be trustable. It has been argued in the literature that one must
excise the supergravity background, leaving a constant $\gamma$ inside. Here we can \textit{prove} that this is indeed the
way the non-perturbative dynamics of the gauge theory translate onto the string side:

When we use that $|z|<\rho_1\implies |T_r(z)|\gg |\alpha_2|$ in this region (recall we are at large $M$), equation
\eqref{eqn:easy} becomes
\begin{equation}
2\pi i\,\gamma = - i\pi \, .
\label{eqn:jump}
\end{equation}
This corresponds to the result we would expect from an excision procedure, where the profile of $\gamma$
is frozen in the excised region to a constant value.
Notice however that $\gamma$ inside the enhançon is not its value at the enhançon radius as is often implied in the supergravity literature, but its real part jumps discontinuously in the large $M$ limit.
This cannot be seen of course in the supergravity analysis and is a purely stringy effect.
The jump of the real part of $\gamma$ has already been observed in \cite{Cremonesi:2009hq}.
\bigskip

Let us recapitulate the lessons learnt in this subsection about the enhançon vacuum. First of all, the $N$ regular
D3-branes just come along for the ride and play absolutely no role. This
just reflects the fact that they do not couple to $\gamma$. Second, the roots of the Seiberg-Witten curve
of the microscopic model enter in the game via the defining equation \eqref{eqn:int.enh}. The analysis of the
branch cuts allows us to discover that while at the first radius with branch cuts, at the scale $|z_0|$, there
are fractional D3-branes present (the branch cuts are exponentially short); at the second radius, at the scale
$\rho_1$, the branch cuts are longer and the brane interpretation no longer holds. Finally, we see that inside this
second region of branch cuts, i.e.\ inside the enhançon, the profile of the supergravity twisted field is constant. This agrees with the
\emph{ad hoc} procedure developed in the supergravity literature to cure the repulson singularity of the background, called
excision. Here everything follows from the quantum properties of the microscopic underlying theory.

\subsection{A generic vacuum}
\label{sec:genvac}

A natural question that comes to mind is how the picture we have obtained for the enhançon vacuum changes when
we consider different vacua. The answer is that, morally, nothing changes. The essential parts of the
discussion above apply as well for different (suitable) choices of $T_r$. Let us be a bit more precise.

Take generic polynomials $T_0$, $T_1$ as in \eqref{T0T1def}; $T_r$ is the quotient $T_0/T_1$. If $T_0$ and $T_1$ share any root, this root factors out of $T_r$ and plays no role.
Indeed, that would indicate the presence of a regular brane, which is invisible to $\gamma$.
Thus, we suppose that we have $M$ pairs of different roots, and we let $r$ of the roots of $T_1$ be zero, corresponding in the perturbative picture to $r$ fractional branes of type 1 sitting at the origin.
Let us look at the branch cuts of the integrand in \eqref{eqn:corr}.
The branch points coming from the first square root solve one of the equations
\begin{equation}
T_r=\frac{\prod_{i=1}^{M}\left(z-\tilde{z}_i\right)}
{\prod_{j=1}^{M}\left(z-z_j\right)}=\pm\alpha_1\,.
\end{equation}
Since $\alpha_1$ is very small, for finite $\tilde{z}_i$ the solutions to this equation are
$z\sim\tilde{z}_k\pm\alpha_1\frac{\prod_{i\neq k}(\tilde{z}_k-\tilde{z}_i)}{\prod_j(z-z_j)}\sim\tilde{z}_k$. The branch
cuts are exponentially short. The story for the branch cuts coming from the second square root is different:
\begin{equation}
T_r=\frac{\prod_{i=1}^{M}\left(z-\tilde{z}_i\right)}
{\prod_{j=1}^{M}\left(z-z_j\right)}=\pm\alpha_2\,.
\end{equation}
Given that $\alpha_2$ is very big, now we will have $M-r$ branch points of the form
$z\sim z_l\pm\alpha_2^{-1}\frac{\prod(\tilde{z}_k-\tilde{z}_i)}{\prod_{j\neq l}(z-z_j)}\sim z_l$, with $l\neq z$ and
short branch cuts associated. Regarding the remaining $r$ branch points, it is easy
to see that they will distribute homogeneously on a ring at the scale
\begin{equation}
|z|\approx \frac{\prod_{i=1}^{M}|\tilde{z}_i|^{\frac{1}{r}}}{\prod_{j=1}^{M-r}|z_j|^{\frac{1}{r}}}|4\qq_1|^{\frac{1}{2r}}=
\left(\frac{2\prod_{i=1}^{M}|\tilde{z}_i|}{\prod_{j=1}^{M-r}|z_j|}\right)^{\frac{1}{r}}\,e^{-\frac{\pi}{2g_s\,r}}\,.
\label{eqn:gen.enh}
\end{equation}
This distribution yields ``large'' branch cuts. 
In order to have a shell of branch cuts that will induce an enhançon mechanism, $r$ must be big.
Since we are taking $g_s$ to be small, we see that only if $r$ is of order $M$, the enhançon
phenomenon will be noticeable. In other words, as we already knew, the enhançon is a large $N$ phenomenon.

In the reasoning above, to obtain the formula \eqref{eqn:gen.enh} we assumed that the $\tilde{z}_i$ and the non-zero $z_j$ were finite,
meaning that they do not vanish in the $\qq_a\to0$ limit. But this condition is actually a bit too restrictive.
The approximations $\tilde{z}_k\pm\alpha_1\frac{\prod_{i\neq k}(\tilde{z}_k-\tilde{z}_i)}{\prod_j(z-z_j)}\sim\tilde{z}_k$
and $z_l\pm\alpha_2^{-1}\frac{\prod(\tilde{z}_k-\tilde{z}_i)}{\prod_{j\neq l}(z-z_j)}\sim z_l$ (by $\sim$ we
mean up to exponentially suppressed corrections $\OO(e^{-(\textrm{sth})M})$) still hold if the roots
$\tilde{z}_i,z_j$ contain $\qq_a$ factors in particular ways. This is the case of the cascading vacuum, where we distribute
$(2K+1)M$ fractional branes on $2K$ shells in order to trigger the baryonic root transitions at the scales where the
perturbative gauge couplings diverge, as discussed after \eqref{eqn:rho_1}. This distribution is characterized by the polynomials
\begin{equation}
T_0=\left(z^M+z_0^M\right)\prod_{i=0}^{K-1}\left(z^{2M}+\qq^{\frac32+2i}z_0^{2M}\right)\,,\qquad
T_1=z^M\prod_{j=0}^{K-1}\left(z^{2M}+\qq^{\frac12+2j}z_0^{2M}\right)\,.
\end{equation}
It was indeed checked in \cite{Benini:2008ir} that for $T_0,T_1$ of the form above, the branch cuts associated to the solutions of $z^{2M}=-\qq^{\frac32+2i}z_0^{2M}$ and $z^{2M}=-\qq^{\frac12+2j}z_0^{2M}$ are exponentially short.

\medskip

In summary, we characterize a point on the Coulomb branch by two monic polynomials, $T_0$ with roots
$\tilde{z}_i$, and $T_1$ with roots $z_j$. We take $M$ of the roots $\tilde{z}_i$ to be the solutions of
$\tilde{z}_i^M=-z_0^M$ (recall this triggers the running below $|z_0|$),
and the rest of the roots to be freely distributed inside the circle of radius $|z_0|$.
When two roots $\tilde{z}_i$ and $z_j$ coincide, this signals the presence of regular branes, about which
we cannot say anything in our approach since they do not couple to $\gamma$. Otherwise, $\gamma(z)$ has small
branch cuts at the positions $\tilde{z}_i,z_j\neq0$. They can be interpreted as fractional branes of type 0 and
type 1 respectively. The roots $z_j=0$ cannot be interpreted as localized fractional branes. When we have a large
number $r$ of them, $r\sim\OO(M)$, the enhançon mechanism takes place at the scale \eqref{eqn:gen.enh}. As can be
seen from \eqref{eqn:gammaSQCD}, $\gamma(z)$ is constant inside the enhançon region (where $|T_r(z)|\ll|\alpha_2|$),
which matches the supergravity excision procedure. Notice that this general analysis does not include
exceptional points, neither the possibility of having roots that scale arbitrarily with $\qq_a$. Although our tools are
general enough to analyse these cases, we have not pursued this direction here.

\section{Outlook}
\label{sec:conc}

As we have already emphasized throughout the paper, our main result is the computation, using field theory
 techniques (although often phrased in a stringy language), of the exact profile of the
twisted supergravity field \eqref{eqn:corr}. From the string point of view, this formula includes all $g_s$ and $\alpha'$
corrections. This allowed us to derive directly from field theory the enhançon mechanism proposed in the supergravity
literature.

Clearly, we have not fully exploited the power of the exact result \eqref{eqn:corr}.
Its validity for any value of $N$ in particular opens the possibility of studying $1/N$ corrections.
These corrections are expected to be important for the exceptional points that are very close to the branch points in the large $N$ limit \cite{Ferrari:2001mg}.
We have not delved either into the physical meaning of the curious ``imaginary'' jump of $\gamma$ at the enhançon radius, noted in \eqref{eqn:jump}.
Since such a jump is not observable in the classical supergravity regime, maybe our techniques could help shed some light on the nature of this stringy effect.
In addition, our results can be generalized in several directions that we believe merit further investigation.
One is the generalization to more general vacua, which fall outside the regime considered in section~\ref{sec:genvac}.
Among them we would like to point out the rather mysterious enhançon bearings of \cite{Benini:2008ir}, where some roots of the polynomial $T_0$ are put at a radius which sits inside the enhançon.
Another possibility is the extension of our findings to generic ADE orbifold singularities. We expect the field theory part of the computation to involve the same techniques we have used, albeit with more complicated integrals to evaluate; the physics should be richer since a more intrincate enhançon mechanism is expected with several enhançon radii.
Although the supergravity solutions have been discovered long ago \cite{Billo:2001vg}, as far as we know, the dual field theories have not been as explored in the literature as that of their $\CC^2/\ZZ_2$ counterpart.

\section*{Acknowledgements}

We would like to thank Riccardo Argurio, Cyril Closset,
Stéphane Detournay, Frank Ferrari, Davide Forcella, Simone Giacomelli, Alberto Lerda,
Alfonso V. Ramallo and Diego Redigolo for enriching discussions.
We are especially grateful to Riccardo Argurio and Stefano Cremonesi for comments on the manuscript.
This work is supported in part by the Belgian Fonds de la Recherche
Fondamentale Collective (grant 2.4655.07), the Belgian Institut
Interuniversitaire des Sciences Nucl\'eaires (grants 4.4511.06 and 4.4514.08),
by the ``Communaut\'e Fran\c{c}aise de Belgique'' through the ARC program
and by the ERC through the ``SyDuGraM'' Advanced Grant. We would also like to acknowledge support
from the ``Mohemian Gravity Brotherhood'' during the last stages of the project.
M.M.\ is a Research Fellow of the Belgian Fonds de la Recherche Scientifique - FNRS.

\appendix

\section{Non-perturbative computation of the correlator}
\label{sec:appendix}

In this appendix we are going to derive the expression \eqref{eqn:corr} for the twisted supergravity field in terms of the two polynomials $T_0$ and $T_1$ specifying the Coulomb branch vacuum.
As explained in the main text, this requires to compute the correlator \eqref{eqn:g.corr}. 
This will be achieved by exploiting the recent results of \cite{Fucito:2012xc,Nekrasov:2012xe}, generalizing to $\nn=2$ quivers the microscopic approach to Seiberg-Witten theory \cite{Nekrasov:2002qd,Nekrasov:2003rj}.
We briefly review the main results of this work and explain in detail how they allow us to derive an explicit expression for the supergravity twisted field.

The main objects of study are the correlators
\begin{equation}
  y_a(z)=\exp \langle \tr \log (z-Z_a) \rangle \, ,
  \label{eqn:yadef}
\end{equation}
where the $Z_a, a=0,1$, are the adjoint fields (normalized to have dimension of length) of the two gauge groups, taken to be $\SU(N+M)_a$.
The functions $y_a(z)$ are generating functions for all correlators of the theory on the Coulomb branch as can be seen by Taylor expanding the logarithm in \eqref{eqn:yadef}
\begin{equation}
  y_a(z)=z^{N+M}\exp \left[-\sum_{k=2}^\infty \frac{1}{k z^k}\langle \tr Z_a^k \rangle \right] \, .
  \label{eqn:yaexp}
\end{equation}
There is no $k=1$ term in this expansion because we are dealing with $\SU(N+M)$ adjoint fields.
If the $Z_a$ were ordinary finite dimensional matrices instead of quantum fields, the Cayley-Hamilton theorem would express $\tr Z_a^k$ for $k> N+M$ in terms of the traces for $k=2,\ldots,N+M$.
These identities satisfied by the traces would ensure that all terms containing negative powers of $z$ in the expansion \eqref{eqn:yaexp} actually vanish and that $y_a$ is a polynomial, the characteristic polynomial of the matrix $Z_a$.
However, this needs not be the case in a quantum field theory\footnote{We thank F. Ferrari for clarifying this point to us.} as product of operators are not \emph{a priori} defined but require a choice of regularization scheme.
This regularization scheme will in general spoil the Cayley-Hamilton identities between the traces resulting in non-polynomial $y_a$.
In particular, this is the case for the most natural regularization scheme for instanton computations in $\nn=2$ theories obtained by turning on a non-commutative deformation and the $\Omega$-background.
Even though the $y_a$ are not polynomials, the main result of \cite{Fucito:2012xc,Nekrasov:2012xe} particularized to the case of the $\CC^2/\ZZ_2$ orbifold is that one can construct two functions of $y_0$ and $y_1$ which are actually polynomials of degree $N+M$, $\tilde T_0$ and $\tilde T_1$.
They read
\begin{align}
  \tilde T_0(z)&=\frac{y_0(z)}{\phi(\qq)}\thet3\left(\frac{y_1(z)^2}{\qq_1 y_0(z)^2} ; \qq^2 \right) \, , \label{eqn:tT0def} \\
  \tilde T_1(z)&=\left( \frac{\qq_1}{\qq_0} \right)^\frac{1}{4}\frac{y_0(z)}{\phi(\qq)}\thet2\left(\frac{y_1(z)^2}{\qq_1 y_0(z)^2} ; \qq^2 \right) \, , \label{eqn:tT1def}
\end{align}
where $\qq=\qq_0\qq_1$ and $\qq_a=e^{2\pi i\tau_a}$ are defined by the two holomorphic gauge couplings of the conformal theory.
We can then write down the following expansions of the $\qq$-Pochhamer symbol $\phi$ and the Jacobi $\theta$-functions,
\begin{align}
  \phi(\qq) &= \prod_{k=1}^\infty (1-\qq^k) \label{eqn:phidef} \, , \\
  \thet2(t; \qq) &= \sum_{n \in \ZZ+\frac{1}{2}} t^n \qq^{\frac{1}{2}n^2} \, , \label{eqn:th2def} \\
  \thet3(t; \qq) &= \sum_{n \in \ZZ} t^n \qq^{\frac{1}{2}n^2} \label{eqn:th3def} \, .
\end{align}
The polynomials $\tilde T_0$ and $\tilde T_1$ are not quite the same as the polynomials $T_0$ and $T_1$ used in the main text since they are not monic, i.e.\ the coefficients of the $z^{N+M}$ terms are not $1$, but rather
\begin{align}
  \tilde T_{0,0} &= \frac{1}{\phi(\qq)}\thet3\left(\frac{1}{\qq_1};\qq^2\right) \, , \label{eqn:T00} \\
  \tilde T_{1,0} &= \left( \frac{\qq_1}{\qq_0} \right)^{\frac{1}{4}}\frac{1}{\phi(\qq)}\thet2\left(\frac{1}{\qq_1};\qq^2\right) \, . \label{eqn:T10} 
\end{align}
Hence, we define the monic polynomials $T_0$ and $T_1$ by
\begin{equation}
  T_0(z)=\frac{\tilde T_0(z)}{\tilde T_{0,0}} \, , \quad T_1(z)=\frac{\tilde T_1(z)}{\tilde T_{1,0}} \, ,
  \label{eqn:defT0T1}
\end{equation}
which coincide with the polynomials used in the main text.

Plugging \eqref{eqn:yadef} into \eqref{eqn:g.corr}, we can express $\gamma$ as
\begin{equation}
  e^{2\pi i\gamma(z)}=\qq_1\frac{y_0(z)^2}{y_1(z)^2} \, .
  \label{eqn:gammay}
\end{equation}
The \RHS\ is the inverse of the argument of the $\theta$-functions in \eqref{eqn:tT0def}, \eqref{eqn:tT1def} and we thus need to invert those relations to obtain $\gamma$ in terms of $\tilde T_0$ and $\tilde T_1$.
For this, we use the following properties of the $\theta$-functions, which can easily be derived from the Fourier expansions \eqref{eqn:th2def} and \eqref{eqn:th3def}.
\begin{enumerate}
  \item \label{prop:1} The functions $\thet2(t;\qq)$ and $\thet3(t;\qq)$ are elliptic, i.e.\ holomorphic functions in $t$ associated to an elliptic curve $\mathcal E$ with complex structure $\qq=e^{2\pi i \tau}$.
    Defining also $t=e^{2\pi i u}$, this elliptic curve is the complex torus $\mathcal E=\CC/\Lambda$ where $\Lambda$ is the lattice $\Lambda=\{u \in \CC \vert u=m+n\tau , (m,n) \in \ZZ^2\}$.
    Holomorphicity of $\thet2$ and $\thet3$ is a consequence of the convergence of the series \eqref{eqn:th2def}, \eqref{eqn:th3def} for $|\qq|<1$ or equivalently $\im\tau>0$.

  \item \label{prop:2} The functions $\thet2(t;\qq)$ and $\thet3(t;\qq)$ enjoy periodicity properties: they are periodic under $u\to u+1$, being functions of $t$ only, and quasi-periodic under $u\to u+\tau$ or equivalently $t\to t\qq$,
    \begin{equation}
      \thet2(t\qq;\qq)= t^{-\frac{1}{2}}\qq^{-1}\thet2(t;\qq) \, , \quad   \thet3(t\qq;\qq)= t^{-\frac{1}{2}}\qq^{-1}\thet3(t;\qq) \, .
      \label{eqn:thperiod}
    \end{equation}

  \item \label{prop:3} The functions $\thet2(t;\qq)$ and $\thet3(t;\qq)$ each have a single simple zero on $\mathcal E$, 
    \begin{equation}
      \thet2(-1;\qq)=0 \, , \quad \thet3(-\qq^{\frac{1}{2}};\qq)=0 \, .
      \label{eqn:thzeroes}
    \end{equation}
\end{enumerate}
To invert \eqref{eqn:tT0def} and \eqref{eqn:tT1def}, we adopt the same strategy as \cite{Nekrasov:2012xe} and define
\begin{align}
  t^2&= \frac{y_1^2}{\qq_1 y_0^2} = e^{-2 \pi i \gamma} \, , \label{eqn:deft} \\
  T_r &= \frac{T_0}{T_1} = \frac{\tilde T_{1,0}}{\tilde T_{0,0}} \left( \frac{\qq_0}{\qq_1}\right)^{\frac{1}{4}} \frac{\thet3(t^2;\qq^2)}{\thet2(t^2;\qq^2)} \label{eqn:defTr} \, .
\end{align}
We can now use the three properties of the $\theta$-functions stated above to derive the following properties of $T_r$.
By property~\ref{prop:1}, $T_r$ is a meromorphic function.
By property~\ref{prop:2}, it is well-defined on $\mathcal E$ because the coefficients in the periodicity relations \eqref{eqn:thperiod} cancel in the ratio \eqref{eqn:defTr}.
By property~\ref{prop:3}, it has two simple poles at $t=\pm i$ and two simple zeroes at $t=\pm i\qq^{\frac{1}{2}}$.
Finally, it is an even function of $u$ since it is evaluated for $t^2$.
The fact that $T_r$ is an even meromorphic function on $\mathcal E$ with prescribed poles and zeroes allows us to rewrite it in a different way.
Indeed, the field of meromorphic functions on an elliptic curve is the field of fractions generated by the two elements $(X(t;\qq),Y(t;\qq))$ (subject to the relation \eqref{eqn:Weqn} to be discussed shortly), where
\begin{align}
  X(t;\qq)&=\wp(u; \tau) 
  \label{eqn:defX} \, , \\
  Y(t;\qq)&=2\pi i t \frac{\d X}{\d t} (t;\qq) = \wp'(u; \tau) \label{eqn:defY}
\end{align}
are the Weierstrass $\wp$-function and its derivative written in the more convenient $(t=e^{2\pi i u},\qq=e^{2 \pi i \tau})$ variables.
The function $\wp$ is an even meromorphic function on $\mathcal E$ with a double pole at the origin.
This implies that $\wp'$ is odd, and hence that $T_r$ can be written as a function of $\wp$ (or equivalently of $X$) only.
To match the poles and zeroes of \eqref{eqn:defTr}, the right combination is
\begin{equation}
  T_r = T_r^\infty \frac{X(t;\qq) - X_0}{X(t;\qq) - X_1} \, ,
  \label{eqn:TrX}
\end{equation}
where
\begin{align}
  X_0=X(i\qq^{\frac{1}{2}};\qq) \, , \quad X_1=X(i;\qq) \, , \quad T_r^\infty=\frac{\tilde T_{1,0}}{\tilde T_{0,0}} \left( \frac{\qq_0}{\qq_1}\right)^{\frac{1}{4}} \frac{\thet3(1;\qq^2)}{\thet2(1;\qq^2)}= \frac{\thet2(\qq_1^{-1};\qq^2)}{\thet3(\qq_1^{-1};\qq^2)}\frac{\thet3(1;\qq^2)}{\thet2(1;\qq^2)} \, ,
  \label{eqn:X0X1tTrinf}
\end{align}
which are found by matching the zeroes at $t=\pm i\qq^{\frac{1}{2}}$, the poles at $t=\pm i$ and the value at $t=1$ (which is a pole of $X$) respectively.
We can now solve for $X$ in \eqref{eqn:TrX},
\begin{equation}
  X[T_r(z)](t;\qq)=\frac{T_r(z) X_1(\qq)- T_r^\infty(\qq_0,\qq_1)X_0(\qq)}{ T_r(z)- T_r^\infty(\qq_0,\qq_1)} \, ,
  \label{eqn:XTr}
\end{equation}
where we have spelled out the full parametric dependence of the different quantities involved.
We are now nearing the end of our journey through the land of elliptic functions: the \RHS\ is now $t$-independent and all that remains is to invert the relation between $t$ and $X$ to obtain $\gamma$ from \eqref{eqn:deft}.

This can be done by recalling that $X$ and $Y$ satisfy the following polynomial equation:
\begin{equation}
  Y(t;\qq)^2=4 X(t;\qq)^3 - g_2(\qq) X -g_3(\qq) \, ,
  \label{eqn:Weqn}
\end{equation}
which realizes the elliptic curve $\mathcal E$ as a projective variety inside $\mathbb P^2$.
The function $X(t;\qq) $ admits the Fourier expansion
\begin{equation}
  X(t;\qq)=-4 \pi^2 \left[ \frac{t}{(1-t)^2}+\frac{1}{12}+\sum_{k=1}^\infty k \frac{\qq^k}{1-\qq^k}(t^k + t^{-k} -2) \right] \, ,
  \label{eqn:Xseries}
\end{equation}
which coincides with the Weierstrass $\wp$-function by uniqueness.\footnote{It is meromorphic and periodic in $u$ with a double pole at $u=0$ of residue one; subtracting this pole gives a function which vanishes at zero. These properties define the Weierstrass $\wp$-function uniquely.}
The coefficients $g_2$ and $g_3$ are modular forms of weight 4 and 6 respectively.
Their Fourier expansions read
\begin{align}
  g_2(\qq) &=(-4\pi^2)^2 \left[ \frac{1}{12}+20\sum_{k=1}^\infty k^3 \frac{\qq^k}{1-\qq^k} \right] \, ,  \label{eqn:defg2} \\
  g_3(\qq) &=(-4\pi^2)^3\left[ -\frac{1}{216}+\frac{7}{3}\sum_{k=1}^\infty k^5 \frac{\qq^k}{1-\qq^k} \right] \, . \label{eqn:g3def}
\end{align}
The equation \eqref{eqn:Weqn} can be proven by showing that $Y^2-4X^3 + g_2 X + g_3$ is holomorphic and hence constant, and showing that this constant vanishes.

The equation \eqref{eqn:Weqn} plays a crucial role.
By plugging the explicit value \eqref{eqn:XTr} into \eqref{eqn:Weqn} and forgetting about the $t$-dependence of $Y$, we obtain the Seiberg-Witten curve of this model.
However, we are not interested in the Seiberg-Witten curve itself but in $t^2$.
It will be convenient to rewrite the equation \eqref{eqn:Weqn} in terms of its three roots $e_i(\qq)$:
\begin{equation}
  Y(t;\qq)^2=4\left[X(t;\qq)-e_1(\qq)\right]\left[X(t;\qq)-e_2(\qq)\right]\left[X(t;\qq)-e_3(\qq)\right] \, ,
  \label{eqn:Weqnroots}
\end{equation}
which are at 
\begin{equation}
  e_1(\qq)=X(-1;\qq), \, \quad e_2(\qq)=X(-\qq^{-\frac{1}{2}};\qq) \, , \quad e_3(\qq)=X(\qq^{\frac{1}{2}};\qq) \, .
  \label{eqn:eroots}
\end{equation}
Combining the equation \eqref{eqn:Weqnroots} with the definition \eqref{eqn:defY} of $Y$ and the value of $X$ in terms of $T_r$ \eqref{eqn:TrX}, we can write
\begin{equation}
   \frac{\d t}{t} = -2\pi i\, \frac{\d X[T_r]}{Y}=-\pi i\,  \frac{T_r'(z)\frac{\d X}{\d T_r}[T_r]}{\sqrt{\prod_{i=1}^3 \left( X[T_r(z)] -e_i \right)}} \, .
  \label{eqn:dt}
\end{equation}
The choice of branch for the square root must be fixed in order to match the perturbative result in the $z\to\infty$ corresponding to the UV of the theory.
The \RHS\ is independent of $t$, hence $t$ can be found by integrating this equation on a contour that does not cross any branch cuts,
\begin{equation}
  \log \frac{t(z)}{t_1}=-\pi i \int_{z^1}^{z} \frac{T_r'(x)\frac{\d X}{\d T_r}[T_r]}{\sqrt{\prod_{i=1}^3 \left( X[T_r(x)] -e_i \right)}} \, .
  \label{eqn:logt1t}
\end{equation}
To fix the lower bound, we use the relation \eqref{eqn:deft} between $t$ and $y_a$.
The large $z$ asymptotics of $y_a$ are $y_a(z)\sim z^{N+M}$ by \eqref{eqn:yaexp}, hence we have
\begin{equation}
  t_1=\lim_{z\to\infty} t(z)=\qq_1^{-\frac{1}{2}} \, .
  \label{eqn:tbound}
\end{equation}
Using the relation between $\gamma$ and $t$ \eqref{eqn:deft}, we finally obtain an explicit expression for $\gamma$,
\begin{equation}
  2\pi i\gamma(z)=2\pi i \tau_1 - 2\pi i\int^{\infty}_{z} \frac{T_r'(x)\frac{\d X}{\d T_r}[T_r]}{\sqrt{\prod_{i=1}^3 \left( X[T_r(x)] -e_i \right)}} \, .
  \label{eqn:gammaexpr}
\end{equation}

The integrand of \eqref{eqn:gammaexpr} can be massaged a bit in order to obtain a simpler expression.
First, one can evaluate $\frac{\d X}{\d T_r}$ from \eqref{eqn:XTr},
\begin{equation}
  \frac{\d X}{\d T_r}[T_r]=\frac{T_r^\infty(X_0-X_1)}{(T_r-T_r^\infty)^2} \, .
  \label{eqn:dXdTr}
\end{equation}
Plugging this into the integrand of \eqref{eqn:gammaexpr} and expanding $X[T_r]$ yields
\begin{equation}
  \frac{T_r'(x)\frac{\d X}{\d T_r}[T_r]}{\sqrt{\prod_{i=1}^3 \left( X[T_r(x)] -e_i \right)}}= - \frac{\beta}{2\pi i}\frac{T_r'(x)}{\sqrt{\prod_{j=0}^3(T_r(x) - E_j)}}\,,
  \label{eqn:intE}
\end{equation}
where
\begin{align}
  \beta &= - 2 \pi i \frac{T_r^\infty}{\sqrt{\prod_{i=1}^3(X_1-e_i)}} \, , \label{eqn:betadef} \\
  E_i &= T_r^\infty \frac{X_0-e_i}{X_1-e_i} \quad \text{for} \quad i=1,2,3 \, , \label{eqn:Eidef} \\
  E_0 &= T_r^\infty \, . \label{eqn:E0def}
\end{align}
Using identities relating the quantities $\wp-e_i$ to the Jacobi $\theta$-functions (see for instance \cite{NIST}), one can prove that $E_0=-E_1$ and $E_2=-E_3$. Hence by defining
\begin{equation}
  \alpha_1=E_2=-E_3 \, , \quad \alpha_2=E_0=-E_1 \, ,
  \label{eqn:alphadef}
\end{equation}
the \RHS\ of \eqref{eqn:intE} can then be further simplified to
\begin{equation}
  \frac{T_r'(x)\frac{\d X}{\d T_r}[T_r]}{\sqrt{\prod_{i=1}^3 \left( X[T_r(x)] -e_i \right)}}= - \frac{\beta}{2\pi i}\frac{T_r'(x)}{\sqrt{(T_r(x)^2 - \alpha_1^2)(T_r(x)^2 - \alpha_2^2)}} \, .
  \label{eqn:intisimple}
\end{equation}
This yields the formula \eqref{eqn:corr} quoted in the main text by plugging \eqref{eqn:intisimple} back into \eqref{eqn:gammaexpr}.

\bibliography{emergentbiblio}
\bibliographystyle{utphys}

\end{document}